\documentclass[twocolumn, preprint]{aastex63}

\usepackage{amsmath,environ}
\usepackage{braket}
\usepackage{listings}
\usepackage{float}
\usepackage{color}
\usepackage{multirow}

\begin{document}

\title{Far-Infrared Line Diagnostics: Improving N/O Abundance Estimates for Dusty Galaxies} 

\author[0000-0002-1605-0032]{B. Peng}
\affiliation{Department of Astronomy, Cornell University, Ithaca, NY 14853, USA}

\author[0000-0003-1874-7498]{C. Lamarche}
\affiliation{Department of Physics and Astronomy, University of Toledo, 2801 West Bancroft Street, Toledo, OH 43606, USA}

\author{G. J. Stacey}
\affiliation{Department of Astronomy, Cornell University, Ithaca, NY 14853, USA}

\author{T. Nikola}
\affiliation{Cornell Center for Astrophysics and Planetary Science, Cornell University, Ithaca, NY 14853, USA}

\author[0000-0002-4444-8929]{A. Vishwas}
\affiliation{Cornell Center for Astrophysics and Planetary Science, Cornell University, Ithaca, NY 14853, USA}

\author[0000-0001-6266-0213]{C. Ferkinhoff}
\affiliation{Department of Physics, Winona State University, Winona, MN 55987, USA}

\author[0000-0002-8513-2971]{C. Rooney}
\affiliation{Department of Astronomy, Cornell University, Ithaca, NY 14853, USA}

\author{C. Ball}
\affiliation{Department of Astronomy, Cornell University, Ithaca, NY 14853, USA}

\author[0000-0002-4795-419X]{D. Brisbin}
\affiliation{Joint ALMA Observatory, Alonso de Cordova 3107, Vitacura, Santiago, Chile}

\author[0000-0003-1397-0586]{J. Higdon}
\affiliation{Department of Physics, Georgia Southern University, Statesboro, GA 30460, USA}

\author[0000-0002-4021-7453]{S. J. U. Higdon}
\affiliation{Department of Physics, Georgia Southern University, Statesboro, GA 30460, USA}

\correspondingauthor{B. Peng}
\email{bp392@cornell.edu}

\begin{abstract}

The Nitrogen-to-Oxygen (N/O) abundance ratio is an important diagnostic of galaxy evolution since the ratio is closely tied to the growth of metallicity and the star formation history in galaxies. 
Estimates for the N/O ratio are traditionally accomplished with optical lines that could suffer from extinction and excitation effects, so the N/O ratio is arguably measured better through far-infrared (far-IR) fine-structure lines. 
Here we show that the [N {\sc iii}]57$\mu$m/[O\,{\sc iii}]52$\mu$m line ratio, denoted $N3O3$, is a physically robust probe of N/O. 
This parameter is insensitive to gas temperature and only weakly dependent on electron density. 
Though it has a dependence on the hardness of the ionizing radiation field, we show that it is well corrected by including the [Ne {\sc iii}]15.5$\mu$m/[Ne {\sc ii}]12.8$\mu$m line ratio. 
We verify the method, and characterize its intrinsic uncertainties by comparing the results to photoionization models.
We then apply our method to a sample of nearby galaxies using new observations obtained with SOFIA/FIFI-LS in combination with available Herschel/PACS data, and the results are compared with optical N/O estimates. 
We find evidence for a systematic offset between the far-IR and optically derived N/O ratio. 
We argue this is likely due to that our far-IR method is biased towards younger and denser H {\sc ii} regions, while the optical methods are biased towards older H {\sc ii} regions as well as diffuse ionized gas. 
This work provides a local template for studies of ISM abundance in the early Universe. 

\end{abstract}


\section{Introduction}
\label{sec:intro}

Since the metal elements heavier than Lithium are formed in stars, their abundance is a key parameter for galaxy evolution studies. 
However, the absolute abundances are particularly difficult to measure, while the relative abundance of Nitrogen-to-Oxygen (N/O) is more reliably obtained. 
The N/O ratio has been shown to be strongly correlated with the metallicity (O/H) and is a probe of star formation history. 
In particular, the N/O ratio follows a segmented relation to metallicity, such that at low metallicities, N/O stays nearly constant, but when $\log$(O/H) $>$ -3.7 it starts to increase with metallicity. 
This trend is seen in H {\sc ii} regions of nearby galaxies \citep[e.g.][]{1990ApJ...363..142G,1993MNRAS.265..199V,1998ApJ...497L...1V,2010ApJ...720.1738P}, large surveys of galaxies \citep{2013A&A...549A..25P} and spatially resolved integral field unit (IFU) observation of galaxies \citep{2016A&A...595A..62P,2017MNRAS.469..151B}. 
This N/O-metallicity relation does not appear to evolve from z=0 up to z=0.4, but the N/O ratio does appear to grow towards lower redshift accompanied by an increase in metallicity \citep{2013A&A...549A..25P}. 
The segmented relationship is thought to result from there being two origins of Nitrogen: the primary production in the supernova events by massive stars, and the secondary production in mostly intermediate mass stars  \citep{1978MNRAS.185P..77E,2000ApJ...541..660H,2003A&A...397..487P}.
When a galaxy is young and the gas is pristine, the primary Nitrogen and Oxygen dominate the chemical reservoir and maintain a near constant N/O ratio; while at later time, the secondary Nitrogen is produced and enters the ISM at a higher rate relative to Oxygen than the primary production, thereby lifting the N/O ratio.
The dual origins of Nitrogen links the N/O ratio to the star formation history of galaxies.
\citet{1978MNRAS.185P..77E} first proposed to use the N/O ratio as an indicator of the `age' of a galaxy, by which we mean the time since the last major star formation event that built up most of the stars in the galaxy. 
This idea is supported by various observations and models including \citet{Unger2000A&A...355..885U}, \citet{2003A&A...397..487P}, \citet{2006MNRAS.372.1069M}, \citet{2016MNRAS.458.3466V}, \citet{2018MNRAS.477...56V}, etc.
In addition to age, N/O can help study other aspects of galaxy evolution like the burstiness of star formation \citep{1990ApJ...363..142G,1999A&A...345..733C,2002A&A...389..106M} and impact of feedback on chemical evolution \citep{2016MNRAS.458.3466V,2017MNRAS.469.3125C}.

The N/O ratio is also vitally important for studying metallicity. 
Directly or indirectly, N/O is integrated into several commonly used metallicity indices. These include the $N2$ parameter \citep[{[}N II{]}/H$\alpha$;][]{2002MNRAS.330...69D}, the $N2O2$ index \citep[{[}N II{]}$\lambda$6584/{[}O II{]}$\lambda$3727;][]{2002ApJS..142...35K}, the $O3N2$ parameter \citep[{[}N II{]}$\lambda$6584/{[}O III{]}$\lambda$5007;][]{2004MNRAS.348L..59P}, the [O\,{\sc iii}]52/[N\,{\sc iii}]57$\mu$m ratio \citep{Unger2000A&A...355..885U,2017MNRAS.470.1218P}, etc. 
These methods actually measure the N/O ratio or the Nitrogen abundance in the first place, then extrapolate to the Oxygen abundance by assuming certain N/O to metallicity relations, regardless of the large scatter in this relationship. 
Moreover, in most of the photoionization model studies of metallicity, the Nitrogen abundance is input as a parameter dependent on the Oxygen abundance by adopting a N/O-metallicity relation. 
The use of a N/O-metallicity relation in the cases above could introduce appreciable uncertainty in those metallicity diagnostics, and systematically undermines the robustness of N/O-metallicity relation calibrated through optical strong line methods \citep{2009MNRAS.398..949P,2019arXiv190604520S,2020ApJ...890L...3S}. 
Therefore it is crucial to obtain precise N/O measurements to better constrain gas phase metallicities and to understand the star formation in galaxies.

Most previous studies of N/O ratio diagnostics focused on the optical strong line ratios, and are based on a simple assumption that the line flux ratio traces the fraction of the amount of emitting ions, which should be equal to the N/O abundance ratio. 
One widely used method is the $N2O2$ parameter \citep{2002ApJS..142...35K}, deemed to measure the N$^+$/O$^+$ ion ratio. 
The $O3N2$ parameter \citep{2004MNRAS.348L..59P} follows a similar logic but relates to the O$^{++}$ ion. 
The lines from another primary element, Sulfur can be used instead of Oxygen lines as in the $N2S2$ parameter \citep[{[}N II{]}$\lambda$6584/{[}S II{]}$\lambda$6717+6731][]{2007MNRAS.381.1719V}. 
These methods are subject to the  common drawbacks of optical lines: optical lines often suffer from dust extinction, especially in dusty galaxies with high star-formation rate (SFR) like Luminous Infrared Galaxies (LIRGs); forbidden line fluxes depend exponentially on the electron temperature,
which in H {\sc ii} regions is comparable to the optical line excitation potentials. 
According to \citet{2009MNRAS.398..949P}, $N2O2$ and $N2S2$ have dispersion of 0.24 and 0.31 dex respectively. 
To cope with these issues, corrections for temperature and density are usually introduced \citep[c.f.][]{1992MNRAS.255..325P,2010ApJ...720.1738P}, though such corrections introduce more statistical errors that could worsen the overall uncertainties.

The far-IR fine-structure lines have significant advantages over optical lines in probing the ISM physical properties. 
At wavelengths much greater than the size-scales of interstellar dust, they are not greatly affected by dust extinction. 
The lines arise from levels only a few hundred K above the ground state, so that they are insensitive to gas temperature in H {\sc ii} regions.
Furthermore, the line emission is typically optically thin, so radiative transfer effects are minor. 
Therefore the far-IR lines of O$^{++}$, N$^+$, and N$^{++}$ ions serve as reliable proxies to diagnose the N/O ratio. 
Far-IR diagnostics of N/O were first introduced by \citet{1983ApJ...271..618L} who used the [N\,{\sc iii}]57$\mu$m/[O\,{\sc iii}]52$\mu$m line ratio, and applied it to a large sample of H {\sc ii} regions in the Milky Way \citep{1987ApJ...320..573L}. 
This diagnostic was later calibrated in \citet{1988ApJ...327..377R} using the photoionization model grids in \citet{1985ApJS...57..349R} to account for the different O$^{++}$ and N$^{++}$ volumes in varied radiation fields. 
This line ratio was explored in the extragalactic context first with ISO observations \citep{Unger2000A&A...355..885U}, then with Herschel/PACS \citep{2011A&A...526A.149N,2017MNRAS.470.1218P} as a probe of metallicity. 
\citet{Unger2000A&A...355..885U} made a correction for ionization based on the [N\,{\sc ii}]122$\mu$m/[N\,{\sc iii}]57$\mu$m line ratio, but the other papers neglected the effect of radiation hardness which has an important effect on the [N\,{\sc iii}]/[O\,{\sc iii}]52 line ratio. 
Also based on Oxygen and Nitrogen far-IR line ratio, [O\,{\sc iii}]88$\mu$m/[N\,{\sc ii}]122$\mu$m is used to estimate metallicity in \citet{2018MNRAS.473...20R}, though it can be argued that the 88/122 um ratio is a better indicator of radiation field hardness \citep{2011ApJ...740L..29F}.
In this paper the [N\,{\sc iii}]/[O\,{\sc iii}]52 line ratio is revisited as a diagnostic of N/O with the effect of the radiation field hardness more carefully considered, and the line ratio probes are calibrated by more recent photoionization model grids that explore a larger parameter space than were given in \citet{1985ApJS...57..349R}. 

This paper is structured as follows: in Sec.~\ref{sec:strong}, the strong line ratio $N3O3$ parameter and  the density corrected $N3O3_{n_e}$ parameter are introduced as first order N/O diagnostics; in Sec.~\ref{sec:model}, the Neon line ratio is introduced to correct for radiation field hardness. 
We then use the photoionization models from the BOND and CALIFA projects to calibrate the relationship between the $N3O3$ parameter and the [Ne {\sc iii}]15.5$\mu$m/[Ne {\sc ii}]12.8$\mu$m line ratio. 
Sec.~\ref{sec:appl} describes the galaxy sample and the reduction of SOFIA/FIFI-LS data on which the $N3O3$ diagnostic is applied. 
The results are then compared with the N/O ratios reported in the literature. 
In Sec.~\ref{sec:summary}, we summarize the main results, and conclude the paper by highlighting the prospects of using the $N3O3$ parameter for the study of chemical abundances and galaxy evolution in high redshift galaxies.

\section{The $N3O3$ Strong Line Method}
\label{sec:strong}

\subsection{The $N3O3$ parameter}
\label{sec:n3o3}

The [N\,{\sc iii}] 57 $\mu$m and [O\,{\sc iii}] 52 $\mu$m lines arise from the ground state term of the N$^{++}$ and O$^{++}$ fine structure configurations. 
Their emitting levels are collisionally excited by electrons in H {\sc ii} regions so that the line ratio is affected by the H {\sc ii} region physical properties and ionization structure in the following way:
\begin{equation}\begin{split}
\label{equ:line_ratio}
	\frac{F_{\mathrm{[N\,III]}}}{F_{\mathrm{[O\,III]52}}} \sim \frac{n(\mathrm{N})}{n(\mathrm{O})}\ \frac{\mathrm{N}^{++}/\mathrm{N}}{\mathrm{O}^{++}/\mathrm{O}}\ \frac{\varepsilon_{\mathrm{[N\,III]}}}{\varepsilon_{\mathrm{[O\,III]52}}}
\end{split}\end{equation}
The first term is the gas phase N/O abundance ratio; the second term is the ratio of the fractions of N and O that are doubly ionized; the last term is the ratio of emissivity, defined here as the power emitted in the line per ion. 
The emissivity is primarily a function of electron density, and more weakly dependent on electron temperature. 
The line emission is assumed to be optically thin.

The [N\,{\sc iii}]/[O\,{\sc iii}]52 line ratio is a good tracer for the N/O abundance ratio for several reasons. First, the energies required to produce both ions are very close (Table~\ref{tab:lines}), so that the  N$^{++}$ and O$^{++}$ ions occupy very similar regions in ISM, typically H {\sc ii} regions of young stars or the ionized regions surrounding Active Galactic Nuclei (AGNs). 
Notice that their formation potentials are just above that of Helium (24.6 eV) which means they share nearly the same volume as He$^+$ \citep{1983ApJ...271..618L}, the ion which dominates the ionization structure in the H {\sc ii} regions with hard radiation field. 
Secondly, their critical densities are very similar so that the [N\,{\sc iii}] to [O\,{\sc iii}] 52 $\mu$m emissivity ratio changes only by a factor of 5 from the low density to high density limit. 
This is much smaller than the variation if the [O\,{\sc iii}] 88 $\mu$m line is used. 
Thus in Equ.~\ref{equ:line_ratio}, the ionization and emissivity ratios nearly cancel out, so that the [N\,{\sc iii}]/[O\,{\sc iii}]52 line ratio is a good first order proxy for the N/O abundance ratio. 
Furthermore, on large scales in galaxies, this doubly ionized line flux probe can be interpreted as arising primarily from a few H {\sc ii} regions with high excitation. 
This is different from lower ionization state lines which come from a much larger collection of H {\sc ii} region excitation environments \citep[e.g. the diffuse ionized ISM; see][]{2017ApJ...846...32D}, and thus are more difficult to interpret.

Our first estimate of the N/O abundance ratio uses only the [N\,{\sc iii}]/[O\,{\sc iii}]52 line flux ratio:
\begin{equation}\begin{split}
\label{equ:n3o3}
    \mathrm{N}/\mathrm{O} \sim N3O3 = \frac{F_{\mathrm{[N\,III]}}}{F_{\mathrm{[O\,III]52}}} \times 0.400
\end{split}\end{equation}
The index is dubbed as $N3O3$ parameter for clarity. The numerical factor is what one obtains in the low density limit ($n_e \ll n_{\mathrm{crit}}$) for Equ.~\ref{equ:line_ratio}. 
For extragalactic work, this is a good approximation since electron densities in starburst galaxies are usually of the order $10^2 \ \mathrm{cm^{-3}}$ \citep{2013ApJ...777..156I}, whereas $n_\mathrm{crit}$ of both lines are above $2 \times 10^3\ \mathrm{cm^{-3}}$. 
We have also assumed an electron temperature of $T_e= 10^4$ K, which is typical for H {\sc ii} regions. 
The exact value chosen has only a small effect on far-IR lines. 
More description of the parameters and details of the calculation can be found in  Appendix~\ref{sec:lowdensity}.

This method is advantageous in theoretical and practical aspects. 
It is based on simple arguments and is more physically robust than other N/O diagnostics, especially those involving the O$^{++}$ to N$^{+}$ ion ratio, since these ions occupy different ionization zones. 
It also benefits from the weak dependence on temperature of far-IR lines over optical methods. 
The simplicity of using only two lines lends great applicability when studying high redshift galaxies, where observation are often difficult or impossible due to telluric absorption, and data is therefore sparse. 
An additional benefit is that as one of the brightest spectral lines from star forming galaxies, the [O\,{\sc iii}] 52 $\mu$m line encodes additional scientific value that is linked to star-formation rates, and that when combined with the equally bright [O\,{\sc iii}] 88 $\mu$m line, the line pair constrains electron density and gas mass.

\subsection{Density correction to $N3O3$ parameter}
\label{sec:n3o3_dens}

Though the low density limit works well in most cases and the simplicity of using only two lines is attractive, it is desirable to correct for the electron density dependence when the density can be estimated, for example by the [O\,{\sc iii}]52/[O\,{\sc iii}]88 ratio or the [N\,{\sc ii}]122/[N\,{\sc ii}]205 ratio. 
Therefore we also provide here a more precise diagnostic tool that is corrected for electron density and temperature. 
The index is denoted $N3O3_{n_e}$ and is defined as
\begin{equation}\begin{split}
\label{equ:n3o3_ne}
    &\mathrm{N}/\mathrm{O} \sim N3O3_{n_e} = N3O3 \times \frac{1 + 0.691 n_e/T_e^{1/2} + 0.0966 n_e^{2}/T_e}{1 + 0.377 n_e/T_e^{1/2} + 0.0205 n_e^{2}/T_e}
\end{split}\end{equation}
where $n_e$ is the electron density value in unit $\mathrm{cm^{-3}}$ and electron temperature $T_e$ in Kelvin. 
The correction factor starts to have a non-negligible effect at $n_e/T_e^{1/2} > 1$, and rises until the value $\sim 4.7$ at the high density end.

Clearly the densities derived from [O\,{\sc iii}]52/88 are the best since it reuses one of the same lines in $N3O3$ thereby measuring densities in the same regions. 
For convenience, we also provide an expression of $N3O3_{n_e}$ that uses the [O\,{\sc iii}]52/88 ratio:
\begin{equation}\begin{split}
\label{equ:n3o3_ne_r21}
    N3O3_{n_e} = N3O3 \times 6.82 \frac{R_{52/88}\left(R_{52/88}+1.01\right)}{2.13+6.26R_{52/88} + R_{52/88}^2}
\end{split}\end{equation}
in which $R_{52/88} = \frac{F_{\mathrm{[O\,III]}52}}{F_{\mathrm{[O\,III]}88}}$ is the [O\,{\sc iii}] line flux ratio.
The detailed derivation of $N3O3$ and $N3O3_{n_e}$ as well as the parameters used for collisional excitation calculation are given in Appendix~\ref{sec:calculation}.

\begin{table}
\centering
\begin{tabular}[c]{lccc}
	\hline
	Line & $E$ & $\lambda$ & $n_\mathrm{crit}$ \\
	 & (eV)& ($\mu$m) & (cm$^{-3}$) \\ \hline
	{[}N\,{\sc iii}{]}$^2\mathrm{P}_{3/2}-^2\mathrm{P}_{1/2}$ & 29.60 & 57.32 & $2.1 \times 10^3$\\
	{[}{O\,{\sc iii}}{]}$^3\mathrm{P}_{2}-^3\mathrm{P}_{1}$ & 35.12 & 51.81 & $3.6 \times 10^3$ \\
	{[}{O\,{\sc iii}}{]}$^3\mathrm{P}_{1}-^3\mathrm{P}_{0}$ & 35.12 & 88.36 & 510 \\
	{[}Ne\,{\sc ii}{]}$^2\mathrm{P}_{1/2}-^2\mathrm{P}_{3/2}$ & 21.56 & 12.81 & $7.0 \times 10^5$ \\
	{[}Ne\,{\sc iii}{]}$^3\mathrm{P}_{1}-^3\mathrm{P}_{2}$ & 40.96 & 15.56 & $2.7 \times 10^5$ \\ \hline
\end{tabular}
\caption{Characteristic of the mid-IR and far-IR lines used for N/O diagnostic. The columns are the energy required to ionize the lower state atom/ion into the listed ion, wavelength, and electron critical density of the spectral line. The data are taken from \citet{2011ITTST...1..241S} and \citet{2016ApJS..226...19F}}
\label{tab:lines}
\end{table}

\section{Photoionization Model Calibration}
\label{sec:model}

\subsection{[Ne\,{\sc iii}]/[Ne\,{\sc ii}] as a radiation hardness tracer}
\label{sec:neonratio}

As stated in the previous section, one of the major benefits of using [O\,{\sc iii}] and [N\,{\sc iii}] is the near co-spatial nature of N$^{++}$ and O$^{++}$ ions in H {\sc ii} regions. 
But in actuality the O$^{++}$ volume is usually smaller than the N$^{++}$ volume: the volumes are only closely matched when the radiation field is hard enough such that the Helium Str{\"o}mgren sphere fills the whole H {\sc ii} region. 
Only in this case can the line ratio trace the abundance ratio at high accuracy without corrections for ionization structure. 
Furthermore, if the radiation field is too hard, our $N3O3$ diagnositic would fail as N$^{++}$ gets ionized into N$^{+++}$ and the [N\,{\sc iii}] flux would decrease. 
Therefore an indicator for the radiation hardness is essential for calibrating the $N3O3$ diagnostic to higher precision.

We use the [Ne\,{\sc iii}] $15.5 \ \mathrm{\mu m}$ to [Ne\,{\sc ii}] $12.8 \ \mathrm{\mu m}$ line ratio as a tracer for radiation hardness. 
Because their ionization potentials are 40.96 eV and 21.56 eV (Table~\ref{tab:lines}), with one higher and the other lower than that of Helium, the Ne$^{++}$ to Ne$^+$ ion ratio closely follows the fraction of volume inside and outside the Helium Str{\"o}mgren sphere in H {\sc ii} region. 
In addition, the critical densities of both lines are $\sim 10^5 \ \mathrm{cm}^{-3}$, much higher than the typical densities in H {\sc ii} regions, so that their emissivity ratio is almost density invariant. 
This is a major advantage over the [N\,{\sc iii}]/[N\,{\sc ii}] ratio, which is another commonly used hardness tracer \citep[c.f.][]{Unger2000A&A...355..885U}: because the [N\,{\sc ii}] lines have relatively low critical densities ($\le 200$ cm$^{-3}$), the line emissivity ratio is much more sensitive to the electron density at $\sim 10^2\ \mathrm{cm}^{-3}$ values. 
Furthermore, since the energy to ionize the third electron off from Neon is 63.45 eV, higher than that energy for Nitrogen (47.4 eV) or Oxygen (54.9 eV), so that the Neon line ratio can still function as hardness tracer in the environments where N$^{++}$ ions are further ionized and the [N\,{\sc iii}]/[N\,{\sc ii}] ratio would fail. 

Other factors that favor the [Ne\,{\sc iii}]15/[Ne\,{\sc ii}]12 ratio as a radiation hardness tracer include that the emitting levels are roughly 1000 K above ground so that they have reduced sensitivity to gas temperature in H {\sc ii} regions compared with optical lines; they lie in the mid-IR range so that they are affected less by extinction, and they lie close in wavelength so the differential extinction correction is small; there is ample data available from mid-IR observatories (Spitzer, ISO, etc); and both Neon lines are often strong in star-forming galaxies. 
But one caveat of the [Ne\,{\sc iii}] line is the possible contamination from AGN.
Though it is not a concern for nearby galaxies, as spectral classification can identify AGNs, and spatial resolution is often enough to separate active nuclei from star-forming regions, AGN contamination might undermine its application on high-redshift objects. 
Fortunately, the nearby [Ne\,{\sc V}] 14.3 $\mu$m line is only bright from AGN on galactic scales and there is a strong correlation between the [Ne\,{\sc iii}] lines and the [Ne\,{\sc V}] line emission in nearby AGN dominated systems, so that the fraction of the [Ne\,{\sc iii}] line that arises from the AGN is well determined by measuring the [Ne\,{\sc V}] line \citep{2007ApJ...655L..73G}.

\subsection{Introduction of photoionization models}
\label{sec:intro_model}

In order to validate the $N3O3$ parameter, and correct it for ionization hardness with the Neon line ratio, we compare our results against photoionization models that account for different stellar populations and include detailed physical calculations on ionization structure and radiative transfer. 
These models can be used to study the uncertainty of this diagnostic as well because they cover a large volume in the parameter space, representative of the diversity of galaxies.

The grids of photoionization models used in this paper are drawn from the CALIFA project \citep{2014A&A...561A.130C} and BOND project \citep{2016MNRAS.460.1739V}. These models are run and hosted by the Mexican Million Models Database \citep[3MDB;][]{Morisset:2015we}, available for public use\footnote{\url{https://sites.google.com/site/mexicanmillionmodels/}}. 
All the models result from running Cloudy photoionization code v17.01 \citep{2017RMxAA..53..385F} through pyCloudy \citep{2013ascl.soft04020M}.

The CALIFA Project is a grid of photoionization models which feature a broad diversity in input stellar populations, with ages ranging from 1 Myr to 14 Gyr. 
The N/O abundance ratio is an independent parameter in the input configuration, spanning from $\log$ N/O = -1.36 to -0.36 in 5 steps. 
These photoionization models were initially run for analyzing IFU observation in the CALIFA survey \citep{2012A&A...538A...8S}.

BOND is a grid of models that also do not assume an N/O-O/H relation, and which cover a broad range in abundance of the two elements as well as radiation field strength. 
These models only use a fixed density of 100 cm$^{-3}$, in either a filled sphere or a thin shell geometry. 
The input starburst ages are 1, 2, 3, 4, 5 and 6 Myr. 
The spectral energy distribution (SED) of the ionizing radiation is obtained from stellar population synthesis models accounting for the appropriate metallicity.

It is essential to use both BOND and CALIFA models, because the former are run on electron density $100 \ \mathrm{cm^{-3}}$ while the latter use $10 \ \mathrm{cm^{-3}}$. 
They complement each other on electron density, which is the parameter that has the largest effect on far-IR fine-structure line emissivity.
Our calibration of the $N3O3$ parameter below justifies the need to combine the two grids, and the usefulness of the density corrected $N3O3_{n_e}$ parameter.

Because BOND and CALIFA cover slightly different regions in the parameter space, only a subset of parameters common for both models is used for consistency. 
The selections of parameters are as follows:
\begin{itemize}
	\item Only filled sphere geometry is used for both grids, corresponding to geometric fraction = 0.03, meaning the inner radius is 3$\%$ that of Str{\"o}mgren sphere radius. Although 3MDB contains results of partial cuts of radiation bounded models, only fully radiation bounded results (H$_{\beta}$ depth $>$ 95\%) are used;
	\item CALIFA has a smaller step size in ionization parameter U than BOND, but covers a slightly smaller range. $\log$U = -4 to -1.5 with step size = 0.5 used for both grids;
	\item BOND only runs on input SEDs with  star burst ages from 1 to 6 Myr with step size 1 Myr, while the CALIFA SEDs range from 1 Myr to 14 Gyr, with a more coarse sampling. Age = 1, 3, 4, 6 Myr are used in BOND, and 1, 3, 4, 5.6 Myr are used in CALIFA in the comparison;
	\item BOND is run on a wider range in $\log$ O/H, but only models with log O/H = -3.2, -3.4, -3.8, -4.0 are used to overlap with CALIFA's range of log O/H = -3.09, -3.31, -3.71, -4.02;
	\item Again only BOND models with $\log$ N/O = -1.25, -1.0, -0.75, -0.5, -0.25 are used, because CALIFA only has log N/O = -1.36, -1.11, -0.86, -0.61, -0.36.
\end{itemize}

\begin{figure*}[ht]
    \centering
    \includegraphics[width = 0.9 \textwidth]{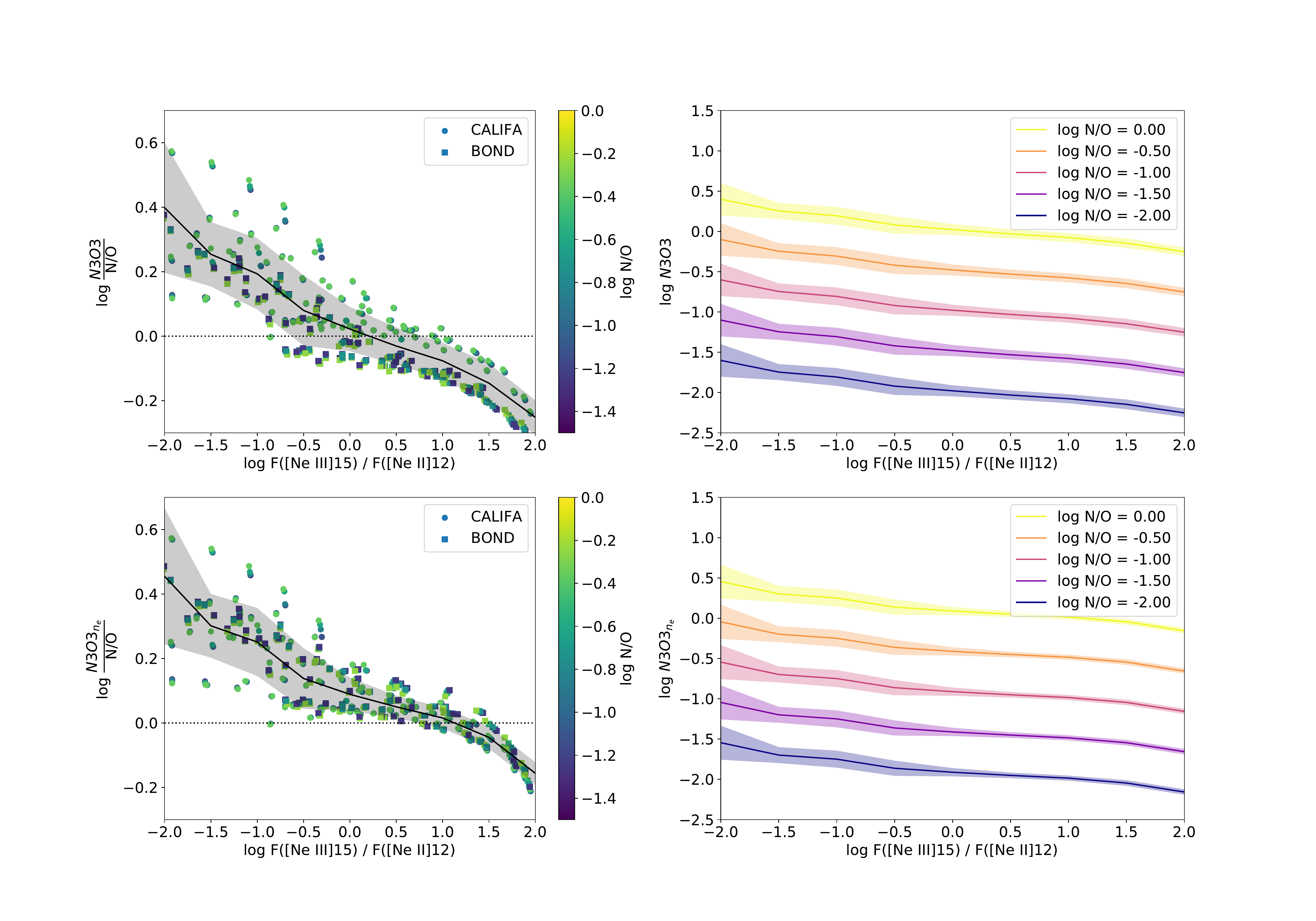}
    \caption{$N3O3$(upper panel) and density corrected $N3O3_{n_e}$(lower panel) parameter of selected photoionization models as a function of the [Ne\,{\sc iii}]15/[Ne\,{\sc ii}]12 ratio. Shown on the left are $N3O3$ and $N3O3_{n_e}$ divided by N/O, in order to show the intrinsic scatter of the calibration; the right panels show the normalized $N3O3$ and $N3O3_{n_e}$ versus Neon line ratio relation as a function of N/O, readily to be applied to estimate the N/O ratio. The shades show the standard deviation in each $\Delta x=0.5$ bin, interpreted as the uncertainty of the relation.}
\label{f:n3o3}
\end{figure*}

These selection cuts result in 480 models in each project, and a combined model count of 960. 
These photoionization models cover the most commonly discussed parameter space in metallicity, the N/O ratio, and the ionization parameter, so that they form a representative collection for calibrating and diagnosing chemical abundance. 
It also samples the relatively low metallicities and young stellar populations that are relevant to the dwarf galaxies and LIRGs tested in this paper, and are usesful for the prospective application to high-redshift counterparts.

\subsection{$N3O3$ calibrated by photoionization models}
\label{sec:n3o3_cal}

The $N3O3$ parameter divided by the N/O ratio of each photoionization model are computed and plotted against the Neon line ratio in the upper left panel in Fig.~\ref{f:n3o3}. 
BOND data points are marked by square and CALIFA as circle symbols. 
All the $N3O3$ are divided and color-coded by the N/O ratio in the models to better illustrate that most of the residual dispersion is due to factors other than the abundance ratio. 
Only data points with $\log$ [Ne\,{\sc iii}]/[Ne\,{\sc ii}] between -2 and 2 are shown in the figure, which reflects  the dynamical range of currently available spectral observations. 
As a result, only 820 out of 960 models are shown in the figure. The mean and standard deviation are calculated in bins of size $\Delta (\log \mathrm{[Ne\,{\sc iii}]/[Ne\,{\sc ii}]})=0.5$, plotted in the black solid lines and enveloping cyan shades. 
The standard deviation in each bin is interpreted as the error of this diagnostic. 
As the [Ne\,{\sc iii}]/[Ne\,{\sc ii}] ratio increases along the x-axis, $\log \ N3O3$ gradually converges and flattens to $0$, the expected value, but it diverges below $0$ at higher Neon ratios. 
This behavior is consistent with the discussion on soft and very hard radiation fields at the beginning of Sec.~\ref{sec:neonratio}. 
It shows that the original $N3O3$ ratio works reasonably well in the regime $-0.5 < \log \mathrm{[Ne\,{\sc iii}]/[Ne\,{\sc ii}]} < 1.5$, and can be calibrated to function in a wider range of radiation field with the help of Neon line ratio.

The data points are color-coded by the N/O ratio, and many points that show a continuous gradient of N/O often cluster together very closely. 
This indicates that as expected, the N/O ratio has a negligible effect on ionizing structure and emissivity, and the scatter seen in the left figure is mainly due to differences in electron density, metallicity and radiation strength, and not due to changes in the N/O ratio. 
This relation is split into abundance groups in the right panels by multiplying with the N/O abundance ratio ranging from $\log$ N/O = 0 to -2. 
This illustrates how the N/O ratio can be determined with the $N3O3$ to Neon line ratio diagram.

One obvious issue with the upper left figure is that the BOND data points settle below CALIFA by about 0.1 dex, especially in the high [Ne\,{\sc iii}]/[Ne\,{\sc ii}] ratio region where $N3O3$ has the highest accuracy. 
This is because BOND is run on electron density $n_e = 100 \ \mathrm{cm^{-3}}$, resulting in lower $\mathrm{\varepsilon_{[N\,III]}}/\mathrm{\varepsilon_{[O\,III]52}}$ ratio.

The calibration using photoionization models validates the effectiveness of $N3O3$, as well as a more precise and reliable relation with an error estimation created with the addition of the Neon line-based radiation hardness tracer. 
However, this calibration also displays some limitations: the $N3O3$ diagnostic works best in hard radiation environments where the $\log$ [Ne\,{\sc iii}]/[Ne\,{\sc ii}] ratio is higher than -0.5 since this is the regime where the N$^{++}$ and O$^{++}$ ionization zones are co-spatial exactly; the $N3O3$ to Neon line ratio relation still has a dispersion of at least 0.03 dex, which largely comes from small variations in the ionization structure caused by metallicity differences; and, the models used for calibration only cover a limited and sparse sample in large parameter space.

\section{Application to a Galaxy sample}
\label{sec:appl}

\subsection{Galaxy sample}
\label{sec:sample}

To demonstrate this new diagnostic, several starburst and low metallicity galaxies in the local universe are selected for application. 
The choice is primarily subject to data and observation scheduling availability, but also various factors including the representativeness of the galaxies and their similarity to the supposedly low-metallicity high redshift galaxies. 
The galaxy sample and their basic properties are summarized in Table~\ref{tab:source}.

\begin{table}[h]
\centering
\begin{tabular}{lll}
\hline
	Galaxy name     & Type        	& Redshift  \\ \hline
	Arp 299         & Interacting LIRG 	& 0.010300  \\
	Haro 3          & BCD    		& 0.003149  \\
	II Zw 40        & BCD    		& 0.002632  \\
	M 83			& Spiral 	  	& 0.001711  \\
	MCG+12-02-001   & LIRG        	& 0.015698  \\
	NGC 2146        & LIRG          & 0.002979  \\
	NGC 4194        & LIRG   		& 0.008342  \\
	NGC 4214        & BCD         	& 0.000970  \\ \hline
\end{tabular}
\caption{Characteristics of galaxy sample. The columns are the common name used in the paper; morphological or luminosity types; redshift of the galaxies. The type information is from NED(NASA/IPAC Extragalactic Database), \citet{2015A&A...578A..53C} and \citet{2009PASP..121..559A}; redshifts are taken from NED.}
\label{tab:source}
\end{table}

Arp 299 is an interacting system comprising two galaxies. 
The nuclear regions of the galaxies are named A (eastern component) and B (southwestern component), separated by about 20". 
The component C located at $\sim$10" north of B is part of the overlapping/external star-forming region.  \citep[see][]{2004ApJ...611..186N}. 
The A source is the brightest IR continuum source in the system and hosts a young starburst estimated to have peaked 6 - 8 Myr ago \citep{2000ApJ...532..845A}. 
B is the second brightest in the far-IR continuum, but interestingly the [O\,{\sc iii}] line coming from B is significantly weaker than its close neighbor C. 
While C shows the opposite feature, its [O\,{\sc iii}] emission is almost as bright as A, but the continuum flux density is an order of magnitude lower than B. 
Because B and C are only separated by $\sim 10$ arcsec and both appear extended, they are blended in both SOFIA/FIFI-LS and Herschel observations, thus they are treated as one source ``B\&C'' in the rest of the paper. 
But the majority of the line emission is actually from component C. 
Both A and B\&C are observed in our SOFIA/FIFI-LS mapping. 
It is in the Great Observatories All-Sky LIRG Survey \citep[GOALS;][]{2009PASP..121..559A} sample, too.

Haro 3 is a Blue Compact Dwarf galaxy (BCD) with two very young starburst regions. 
One of the two regions hosts a starburst younger than 5 Myr, and the other is $8 \sim 10$ Myr old \citep{2004AJ....128..610J}. 
The stellar mass in Haro 3 is $\sim 10^6 \ \mathrm{M_\odot}$, and SFR $\sim 0.8 \ \mathrm{M_\odot / yr}$ measured by combining FUV and MIR photometry \citep{2014A&A...568A..62D}. 
It is included in the Herschel Dwarf Galaxy Survey sample \citep[DGS;][]{2015A&A...578A..53C}.

II Zw 40 is also a Blue Compact Dwarf galaxy with a young starburst. 
The stellar mass is estimated to be $\sim 9 \times 10^{5} \ \mathrm{M_\odot}$, and SFR $\sim 0.8  \ \mathrm{M_\odot /yr} $\citep{2014A&A...568A..62D}. 
The age of the starburst is estimated to be $\sim 3 \ \mathrm{Myr}$  \citep{2018ApJ...865...55L}. 
II Zw 40 is also among the DGS sample.

MCG+12-02-001 is a LIRG and an interacting system. 
The stellar mass is about $8 \times 10^{10} M_\odot$ and SFR is 30 to 55$\ \mathrm{M_\odot/yr}$ \citep{2014A&A...568A..62D,2010ApJ...715..572H}. 
The starburst age is found to be 40–200 Myr from SED fitting \citep{2015A&A...577A..78P}. 
It is in the GOALS sample, but this source has very little optical observational data available.

NGC 2146 is a barred spiral galaxy, and one of the nearest luminous infrared galaxies. The star formation rate is estimated to be 7.9 $\mathrm{M_\odot/yr}$, dominated by the nuclear starburst. It is in the GOALS sample. 
NGC 2146 has Infrared Space Observatory (ISO) detection of the [N\,{\sc iii}] line as well as both of the [O\,{\sc iii}] lines. 
Besides, it is the only source not classified as ``extended,, in \cite{2008ApJS..178..280B} and with all three lines detected, plus it has ample optical and mid-infrared data available to aide our N/O measurement. 
While the star formation activities in NGC 2146 are extended over 2' scales \citep[][]{1991ApJ...373..423S}, the ISO beam is uniform at about 70" for these lines and therefore provides good measurements of the inner 4.9 kpc region of this galaxy. 

NGC 4194 is a LIRG and a minor merger. 
Estimates for its SFR are not in agreement, with SFR ranging from $\sim 46 \ \mathrm{M_\odot /yr}$ by $H_\alpha$ observation \citep{2006AJ....131..282H} to $\sim 13 \ \mathrm{M_\odot /yr}$ in far-IR \citep{2014A&A...568A..62D}. 
Most of the star formation happens in the super star clusters thought to be 5 to 15 Myr old \citep{2014A&A...569A...6K,2007MNRAS.381..228P}. It is in the GOALS samples.

NGC 4214 is a Blue Compact Dwarf galaxy with 2 major components: region I is the larger component centered on the nucleus and is brighter in far-IR; region II is a smaller source located about 30 arcsecond to the southeast. 
Because only region I is within the field of view in our SOFIA/FIFI-LS observations, the discussion in this paper focuses on this component. 
The SFR in NGC 4214 as a whole is $0.063 \ \mathrm{M_\odot /yr}$ \citep{2014A&A...568A..62D}. 
Its stellar population age is estimated at only 4 to 5 Myr through FUV photometry \citep{2007MNRAS.381..228P} and UV point source observations \citep{2011ApJ...735...22W}. 
It is in the DGS sample.

M83 is a nearby spiral galaxy with a prominent bar. At present, only the central regions are mapped in the [O\,{\sc iii}] 52 $\mu$m line by SOFIA/FIFI-LS, so only the nuclear region of M83 is discussed here. 
M83 hosts a circumnuclear starburst, which displays a burst age gradient along the star-forming arc spanning from 6 to 8 Myr \citep{2008MNRAS.385.1110H,2010MNRAS.408..797K}.

\subsection{SOFIA/FIFI-LS data}
\label{sec:sofia}

One major limitation of the application of the far-IR diagnostics we advocate here is the limited availability of the [O\,{\sc iii}]52 and [N\,{\sc iii}] line measurements. 
During its lifetime, the Herschel Space Observatory \citep{2010A&A...518L...1P} provided far-IR spectroscopy at unprecedented sensitivity and spatial resolution. 
However, the 52 $\mu$m [O\,{\sc iii}] line is outside the 55 to 210 $\mu$m spectral range of PACS \citep{2010A&A...518L...2P} for z $<$ 0.06 objects. 
Furthermore the [N\,{\sc iii}] line, which is typically $\sim 5$ times weaker than the [O\,{\sc iii}] line was not often observed so that Herschel measurements of both the [O\,{\sc iii}]52 and [N\,{\sc iii}] lines are very rare. 
Herschel was decommissioned in 2013. 
Luckily NASA's SOFIA observatory \citep{2018JAI.....740011T} is in operation and can provide the spectroscopy we need for the science presented here. 

FIFI-LS \citep{2018JAI.....740003F} is a Far-IR integral field unit onboard SOFIA, with the design and capabilities similar to Herschel/PACS instrument, but with a shorter wavelength spectral cut-off. 
In the 51 - 125 $\ \mathrm{\mu m}$ ``blue channel'', it has $5 \times 5$ spatial pixels (spaxels) of size $6 "/\mathrm{pixel}$. 
Observations are dithered, so that our maps have finer spatial sampling. The velocity resolution is about 300 km/s at 50 - 60 $\mathrm{\mu m}$.

We obtained new far-IR spectroscopic observations for all of our sources except NGC 2146 over a period from February 2017 to May 2019 using FIFI-LS on SOFIA.  
The newly acquired spectral observations are listed in Table~\ref{tab:obs}. 
All the SOFIA data have gone through the level 4 pipeline reduction, which is chopped and corrected for atmospheric transmission. 
However, for NGC 4214 and MCG+12-02-001 the atmospheric transmission around the [N\,{\sc iii}] line is problematic, and to produce the [N\,{\sc iii}] spectra, we manually corrected the raw data for the atmosphere transmission using the information in the spectral cube. 
This reduction shows a line flux similar to the pipeline reduction, but with a better line shape and improved signal-to-noise ratio (SNR). 
For other data sets which were pipeline processed we also manually restored data channels blanked out by the pipeline, since the FIFI-LS pipeline blanks channels with transmission $<$ 0.6. 
We find that this puts too harsh a standard in accepting data, and restoring the blanked channels can improve the baseline determination.

\begin{table}[h]
\tiny
\centering
\begin{tabular}{llll}
	\hline
	Target                               & Line      & Obs-ID                         & T$_\mathrm{exp}$ [s]  \\ \hline
	Arp 299                              & [O\,{\sc iii}]52 & P\_2017-02-28\_FI\_F379B200754 & 2949.12    		   \\
	Haro 3                               & [O\,{\sc iii}]52 & P\_2018-11-07\_FI\_F525B01652  & 1413.12            \\
	II Zw 40                             & [O\,{\sc iii}]52 & P\_2018-11-06\_FI\_F524B01631  & 768.00             \\
	II Zw 40                             & [N\,{\sc iii}]   & P\_2018-11-06\_FI\_F524B01731  & 1536.00            \\
	M83 nucleus						 & [O\,{\sc iii}]52 & P\_2019-05-04\_FI\_F565B400205 & 2119.68			   \\
	MCG+12-02-001                        & [O\,{\sc iii}]52 & P\_2019-05-01\_FI\_F562B100373 & 1320.96            \\
	MCG+12-02-001                        & [N\,{\sc iii}]   & P\_2019-05-01\_FI\_F562B100464 & 1351.68            \\
	NGC 4194                             & [O\,{\sc iii}]52 & P\_2019-05-09\_FI\_F568B00338  & 430.08             \\
	NGC 4194                             & [N\,{\sc iii}]   & P\_2019-05-09\_FI\_F568B00366  & 860.16             \\
	NGC 4214 region I                    & [O\,{\sc iii}]52 & P\_2019-02-28\_FI\_F549B200462 & 1320.96            \\
	NGC 4214 region I                    & [N\,{\sc iii}]   & P\_2019-05-08\_FI\_F567B00512  & 1413.12            \\ \hline       
\end{tabular}
\caption{Table of SOFIA/FIFI-LS observations. The columns are the target of the field; the observed far-IR line; the associated SOFIA observation ID; and the exposure time in seconds.}
\label{tab:obs}
\end{table}

\begin{table*}[ht]
\centering
\begin{tabular}{llllll}
\hline
    Galaxy name        &  {[}N\,{\sc iii}{]} &   {[}O\,{\sc iii}{]}52   & {[}O\,{\sc iii}{]}88     & {[}Ne\,{\sc ii}{]}12     & {[}Ne\,{\sc iii}{]}15 \\ \hline
    Arp 299 A   & $7.3 \pm 0.5 ^a $ & $40.0 \pm 4.28 $    & $28 \pm 0.32 ^a $ & $23.7 \pm 0.26 ^b$    & $5.70 \pm 0.098 ^b$\\
    Arp 299 B\&C   & $7.2 \pm 0.13 ^a $    & $30.8 \pm 3.72 $ & $30 \pm 0.26 ^a $     & $10.4 \pm 0.27 ^b$   & $5.44 \pm 0.098 ^b$ \\
    Haro 3      & $1.23 \pm 0.17 ^c$    & $26.9 \pm 2.99 $   & $18.4 \pm 0.4 ^c $   & $3.52 \pm 0.13 ^c $  & $9.84 \pm 0.74 ^c $\\
    II-Zw 40    & $5.51 \pm 4 $    & $48.6 \pm 4.52$   & $35.9 \pm 0.4 ^c $    & $0.735 \pm 0.079 ^c $ & $14.1 \pm 0.9 ^c $\\
    M83 nucleus     & $16.6 \pm 1.03 ^d $  & $22.7 \pm 3.03 $    & $21.7 \pm 0.70 ^d $   & $50.3 \pm 1.98 ^d $   & $2.93 \pm 0.077 ^d $ \\
    MCG+12-02-001   & $5.29 \pm 1.53 $    & $30.5 \pm 3.09 $    & $23.4 \pm 2.4 ^e $    & $20.1 \pm 0.21 ^b$    & $3.7 \pm 0.067 ^b $\\
    NGC 2146    & $55.1 \pm 5.9 ^e $ & $151.4 \pm 20.1 ^e $ & $157.7 \pm 6.5 ^e $ & $68.2 \pm 0.80 ^b $ & $9.81 \pm 0.123 ^b$ \\
    NGC 4194    & $6.5 \pm 2.2 ^e $ & $31.5 \pm 2.8 $ & $20.6 \pm 1.4 ^e$ & $17.57 \pm 0.14 ^b $  & $5.62 \pm 0.06 ^b $\\
    NGC 4214 region I   & $1.96 \pm 0.70 $    & $17.5 \pm 1.31 $    & $31.9 \pm 0.62 ^f $   & $8.98 \pm 0.22 ^f $   & $18.7 \pm 0.14 ^f $\\ \hline
\end{tabular}
\caption{The flux of [N\,{\sc iii}], [O\,{\sc iii}]52, [O\,{\sc iii}]88, [Ne\,{\sc ii}]12 and [Ne\,{\sc iii}]15 lines used to obtain the N/O abuncance. The unit of flux is $10^{-16} \ \mathrm{W/m^2}$. The line fluxes without superscripts are the SOFIA/FIFI-LS data presented in this work, while the superscripts correspond to the references: (a) \citet{InPrepArp299}; (b) \citet{2013ApJ...777..156I}; (c) \citet{2015A&A...578A..53C}; (d) \citet{2016ApJS..226...19F}; (e) \citet{2008ApJS..178..280B}; (f) \citet{2015A&A...580A.135D}}
\label{tab:flux}
\end{table*}

The SOFIA/FIFI-LS data are delivered as spectral cubes, and the process of measuring the spectra is as follows: first, the raw data is manually corrected for transmission to fill in the missing channels in the pipeline corrected data; second, weights ($w$) are calculated as $w = \mathrm{N} \times \tau$ for each spaxel, where $N$ is the number of scans in each observation and $\tau$ is the transmission.  
The error is calculated based on the rms in spectral dimension, taking into account that $\sigma \propto 1/\sqrt{w}$. 
The continuum map is then computed at each spatial position as the weighted average of the channels free of the spectral line. 
The continuum map is later subtracted from the whole spectral cube to produce a continuum subtracted cube, which is collapsed in spectral dimension to produce a line moment 0 map. 
Now an ellipse is defined as the emitting region of the spectral line, and integrated in each channel to get the spectrum. 
The error estimation takes into account that the noise is correlated across the $5.5"$ beam. 
Finally the line channels in the spectrum are added together as the line flux measurement.

The integrated line flux map and spectra for all but NGC 4194 [N\,{\sc iii}] data are presented in Appendix~\ref{sec:sofia_data}. 
The NGC 4194 [N\,{\sc iii}] observation is not used in this paper, because it lies right beside a deep and wide telluric feature, and another narrow feature is at +200 km/s with respect to the expected source line center. 
The ISO [N\,{\sc iii}] flux is used instead. 
For each source we include the line flux map and the continuum subtracted spectrum. 
On the line map we plot an ellipse showing the region from which the spectrum is extracted.

There are several caveats on the spectra. 
Instead of matching the same areas across different lines, we choose to pick the integration region such that it encloses most of the emission. 
This is justified by two reasons: first it would give us the highest SNR for our flux measurement, which is essential for [N\,{\sc iii}] observations that in some cases only have tentative detections; second, the Herschel/PACS line fluxes used in this work are mostly integrated over a 5 by 5 map, so maximizing the flux in SOFIA/FIFI-LS observation ensures to include any extended emission contained in the PACS data. 
This strategy works well when we compare the SOFIA extracted fluxes with available ISO observations which has an even larger beam so that it could include more extended emission, and find they agree within $1 \sigma$ error. 
Other SOFIA data, especially the [N\,{\sc iii}] observations, also suffer from the low and/or highly variable transmission. In addition, many spectra have excess noise and sometimes systematic trends in their long and/or short wavelength edge channels. 
This is because FIFI-LS can saturate and enter a non-linear regime when the foreground (sky + telescope) emission is too strong, and the integration times on the edge channels are much less than those on the central channels as the instantaneous FIFI-LS bandwidth is swept across the line to increase spectral coverage (private communication with the SOFIA help desk). 
These effects complicate baseline determination. 
The [N\,{\sc iii}] line from II Zw 40 is also very close to a deep absorption feature. 
Its baseline turns downwards at $v_\mathrm{rest} > 150 \ \mathrm{km/s}$ and the error rapidly increases as the telluric transmission declines, making the absolute line flux measurement unreliable. 
Though the line channels add up to flux = $5.51 \pm 1.76 \times 10^{-16} \ \mathrm{W/m^2}$, the asymmetric line shape indicates that up to $30\%$ of the flux could be missing. 
Thus we assess the uncertainty of the measurement should be $4 \times 10^{-16}  \ \mathrm{W/m^2}$, which is used in the paper.

\subsection{Ancillary data}
\label{sec:other_data}

Herschel/PACS \citep{2010A&A...518L...2P} provides the largest archive of [N\,{\sc iii}] and [O\,{\sc iii}] spectral observations to date. 
For the sample galaxies, all but NGC 4194 and MCG+12-02-001 have [N\,{\sc iii}] spectra taken by PACS. 
Most of the data used here are already processed in \citet{2015A&A...578A..53C} and \citet{2016ApJS..226...19F}. 
Because PACS has a similar sized of field of view as our FIFI-LS maps, the flux measurements by FIFI-LS secure any loss of extended emission compared with that of PACS.

Infrared Space Observatory \citep[][]{1996A&A...315L..27K} supplies the vital far-IR data for MCG+12-02-001, NGC 2146 and NGC 4194. 
The data were obtained with the ISO/LWS \citep{1996A&A...315L..38C} and are spatially unresolved. 
The line fluxes are taken from \citet{2008ApJS..178..280B}, and the explanation of data reduction details can be found in that paper.

The mid-IR [Ne\,{\sc ii}] 12.8 $\mu$m and [Ne\,{\sc iii}] 15.5 $\mu$m lines are originally observed with Spitzer/IRS in high-resolution mode \citep{2004ApJS..154....1W,2004ApJS..154...18H}. 
The flux data are directly taken from \citet{2013ApJ...777..156I} and \citet{2015A&A...578A..53C}, where a description of data processing can be found.

\begin{figure*}[!htbp]
    \centering
    \includegraphics[width = \textwidth]{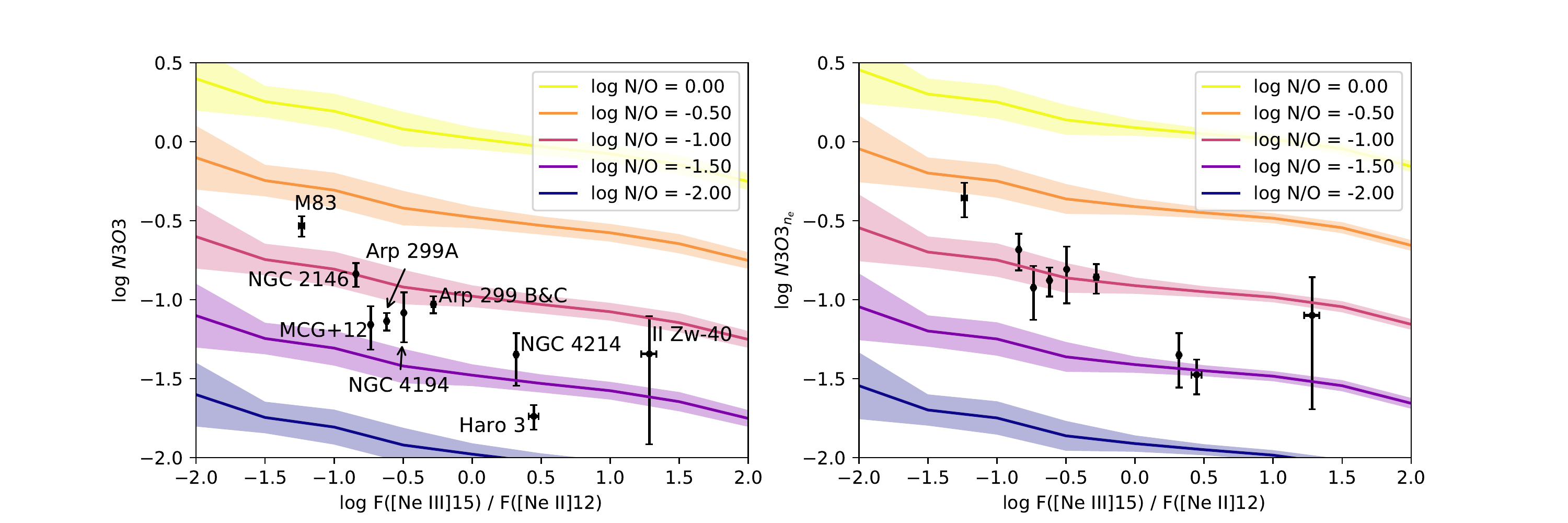}
    \caption{The original and density corrected $N3O3$ of sample galaxies are plotted in the left and right figures respectively, with model calibration lines of $N3O3$ and $N3O3_{n_e}$ shown in the background and color-coded by N/O = -2.0 to 0. The relative position between data point and calibrated diagnostic lines indicates the N/O ratio of each galaxy. Each data point is labeled with its name in the left panels, which can be mapped to the right panel by its position on the x-axis.}
\label{f:result}
\end{figure*}

\begin{table*}[!htbp]
\centering
\begin{tabular}{lcclccc}
\hline
    Galaxy name & \multicolumn{2}{c}{Strong line method} & \ & \multicolumn{2}{c}{Model calibration} & $\log$ {[}Ne\,{\sc iii}{]}/{[}Ne\,{\sc ii}{]} \\ \cline{2-3} \cline{5-6}
     & $\log N3O3$ & $\log N3O3_{n_e}$ & & $\log$ N/O by $N3O3$ & $\log$ N/O by $N3O3_{n_e}$ &  \\
    (1) & (2) & (3) & & (4) & (5) & (6) \\ \hline
    Arp 299 A & $-1.14^{+0.052}_{-0.059}$ & $-0.88^{+0.082}_{-0.101}$ & & $-1.24^{+0.119}_{-0.164}$ & $-1.04^{+0.122}_{-0.171}$ & -0.62 \\
    Arp 299 B\&C & $-1.03^{+0.050}_{-0.057}$ & $-0.86^{+0.083}_{-0.103}$ & & $-1.08^{+0.102}_{-0.133}$ & $-0.97^{+0.109}_{-0.145}$ & -0.28 \\
    Haro 3 & $-1.74^{+0.071}_{-0.085}$ & $-1.48^{+0.096}_{-0.124}$ & & $-1.71^{+0.089}_{-0.113}$ & $-1.53^{+0.101}_{-0.133}$ & 0.45 \\
    II Zw 40 & $-1.34^{+0.239}_{-0.572}$ & $-1.10^{+0.242}_{-0.594}$ & & $-1.23^{+0.242}_{-0.595}$ & $-1.08^{+0.243}_{-0.602}$ & 1.28 \\
    M83 nucleus & $-0.53^{+0.060}_{-0.069}$ & $-0.36^{+0.096}_{-0.123}$ & & $-0.75^{+0.118}_{-0.163}$ & $-0.63^{+0.134}_{-0.195}$ & -1.23 \\
    MCG+12-02-001 & $-1.16^{+0.116}_{-0.159}$ & $-0.92^{+0.138}_{-0.204}$ & & $-1.29^{+0.152}_{-0.237}$ & $-1.12^{+0.163}_{-0.264}$ & -0.73 \\
    NGC 2146 & $-0.84^{+0.068}_{-0.081}$ & $-0.68^{+0.100}_{-0.131}$ & & $-0.99^{+0.126}_{-0.177}$ & $-0.90^{+0.137}_{-0.201}$ & -0.84 \\
    NGC 4194 & $-1.08^{+0.130}_{-0.187}$ & $-0.81^{+0.143}_{-0.215}$ & & $-1.16^{+0.162}_{-0.261}$ & $-0.95^{+0.164}_{-0.267}$ & -0.49 \\
    NGC 4214 region I & $-1.35^{+0.135}_{-0.197}$ & $-1.35^{+0.139}_{-0.206}$ & & $-1.34^{+0.145}_{-0.219}$ & $-1.41^{+0.143}_{-0.215}$ & 0.32 \\ \hline
\end{tabular}
\caption{N/O ratio derived through the $N3O3$ strong line methods and model calibrations. Columns are: (1) name of galaxy as in Table~\ref{tab:source}; (2) $N3O3$ parameter calculated for each galaxy; (3) density corrected $N3O3_{n_e}$ parameter; (4) N/O ratio derived from $N3O3$ parameter photoionization model calibration, as in the left panel of Fig.~\ref{f:result}; (5) N/O from density corrected $N3O3_{n_e}$ parameter as in the right panel of Fig.~\ref{f:result}; (6) [Ne\,{\sc iii}]15/[Ne\,{\sc ii}]12 line flux ratio used for model calibration.}
\label{tab:result}
\end{table*}

Arp 299 and NGC 4214 are two exceptions as they contain multiple components. 
The far-IR data of Arp 299 are from \citet{InPrepArp299}, where fluxes are measured for A and B\&C components separately within a Herschel/PACS map. 
For NGC 4214, we use both the far-IR and the mid-IR line flux values reported in \citet{2015A&A...580A.135D} which measures two regions individually.

In addition to the Neon lines used for calibration, all sources except NGC 4214 have [Ne\,{\sc V}] 14.3 $\mu$m flux upper limits reported in the same references. The upper limits are at least one order of magnitude lower that their [Ne\,{\sc iii}] 15.5 $\mu$m fluxes, indicating negligible AGN contribution. For NGC 4214, the resolved photometric study in \citet{2011ApJ...735...22W} and PDR modeling in \citet{2015A&A...580A.135D} show the central region is dominated by starburst activity with no trace of AGN. Thus all the sources in our sample have little to none AGN contamination in their [Ne\,{\sc iii}]15 and [Ne\,{\sc ii}]12 line fluxes, and the photoionization model calibration based on stellar population synthesis is applicable to this sample.

All the data we use in our application of the $N3O3$ diagnostic are summarized in Table~\ref{tab:flux}.

\subsection{Results}
\label{sec:result}

The $N3O3$ parameter is calculated for each galaxy and then corrected for density by using the [O\,{\sc iii}]52/88 line ratio to get $N3O3_{n_e}$. 
With these strong line parameters in hand, we computed the [Ne\,{\sc iii}]15/[Ne\,{\sc ii}]12 line ratio, and compared it with the calibrated diagnostic as in Sec.~\ref{sec:n3o3_cal} to derive the high-precision N/O abundance ratio. 
The original and density corrected $N3O3$ parameters are plotted against our model calibration in Fig.~\ref{f:result}. 
Their positions in the diagram represent the N/O ratio estimates. 
The values of strong line parameters and the calibrated N/O ratio are listed in Table~\ref{tab:result}.

\begin{table*}[ht]
\centering
\begin{tabular}{lll}
\hline
	Galaxy & Optical $\log$ N/O & Far-IR $\log$ N/O \\ \hline
	Arp 299 A & -0.85$^{+0.026}_{-0.028}$ $^a$ & $-1.04^{+0.122}_{-0.171}$ \\
	Arp 299 B\&C & -0.71$^{+0.026}_{-0.028}$ $^a$ & $-0.97^{+0.109}_{-0.145}$ \\
	Haro 3 & -1.13$^{+0.031}_{-0.033}$ $^a$, -1.06$^{+0.088}_{-0.107}$ $^b$, -1.29 $^c$, -1.35 $^d$,  & $-1.53^{+0.101}_{-0.133}$ \\
	II Zw 40 & -1.30$^{+0.029}_{-0.031}$ $^a$, -1.44 $^c$, -1.44 $^d$, -1.052$^{+0.059}_{-0.077}$ $^e$ & $-1.08^{+0.243}_{-0.602}$ \\
	M83 & -0.63$^{+0.028}_{-0.042}$ $^b$ & $-0.63^{+0.134}_{-0.195}$ \\
	NGC 2146 & -0.77$^{+0.029}_{-0.031}$ $^a$, -1.06$^{+0.049}_{-0.059}$ $^b$ & $-0.90^{+0.137}_{-0.201}$\\
	NGC 4194 & -0.59$^{+0.026}_{-0.028}$ $^a$, -0.5 $^c$ & $-0.95^{+0.164}_{-0.267}$ \\
	NGC 4214 region I & -1.30$^{+0.029}_{-0.031}$ $^a$, -1.28$^{+0.017}_{-0.018}$ $^b$, -1.30 $^{d,e}$,  & $-1.41^{+0.143}_{-0.215}$ \\ \hline
\end{tabular}
\caption{Comparison of the N/O abundance ratio derived by optical methods and that computed by $N3O3_{n_e}$ to Ne line ratio model calibration (Column (5) in Table~\ref{tab:result}). The superscripts correspond to (a) N/O calculated by $N2S2$ index, using spectroscopic data from \cite{2006ApJS..164...81M}; (b) N/O reported in \cite{2019A&A...623A...5D}; (c) \cite{2005A&A...437..849S}; (d) \cite{2019A&A...626A..23C}; (e) \cite{1996ApJ...471..211K}}
\label{tab:compare}
\end{table*}

The derived N/O ratio covers a large range from -1.6 to -0.5 in Table~\ref{tab:result}. 
The N/O of different types of galaxies also cluster together: those of dwarf galaxies are systematically lower than LIRGs, and M83 nucleus has the highest value. 
This is consistent with our understanding of the N/O evolution along with the starburst age and metallicity. 
The errors also change in an increasing trend as more spectral lines are used and more statistical uncertainty is introduced, but the systematic uncertainty in the calibration would decrease. 
And the final error budget is often dominated by the [O\,{\sc iii}]52 and [N\,{\sc iii}]57 line flux errors.

Notice that in the case where the noise is dominated by these two lines, the strong line methods are good estimations of the N/O abundance ratio. 
Comparing the $N3O3$ in Column (2), the most simplistic estimation of N/O using only two spectral lines, to the value in Column (5) which is the most precise and reliable diagnostic utilizing 5 lines, the value changes by $< 0.15$ dex for all targets except Haro 3 and II Zw 40. 
This small change is because the effect of density correction which would lift N/O ratio result, is offset by the model calibration for the $\log $[Ne\,{\sc iii}]/[Ne\,{\sc ii}] in the -0.5 to 1.5 range which would decrease the N/O estimate. 
For the exceptions Haro 3 and II Zw 40, they have moderately high electron densities as well as hard ionization conditions, resulting in noticeable differences between $N3O3$ and the model calibration, but $N3O3_{n_e}$ parameter instead shows great agreement with the model calibrated results. 
This indicates that the $N3O3$ parameter is pretty good an estimator for the N/O abundance when the availability of data is limited, or when the uncertainty is dominated by observational error instead of the intrinsic dispersion of diagnostic, like in the case of samples used here. 
But for galaxies with moderately high electron density and hard radiation field, such as some dwarf galaxies and high-redshift galaxies, $N3O3_{n_e}$ could yield a better N/O estimate than $N3O3$. 

However, using multiple lines and model calibration is still advantageous as it can mitigate the effects of density or radiation hardness in a way that is physically sound. 
The statistical errors may increase with more lines used, but systematic errors will decrease, so that the values obtained with the density and radiation field correcting lines are more reliable. 
Therefore the diagnostic method using more spectral lines is always recommended.

\begin{figure}[h]
	\centering
	\includegraphics[width = 0.5 \textwidth]{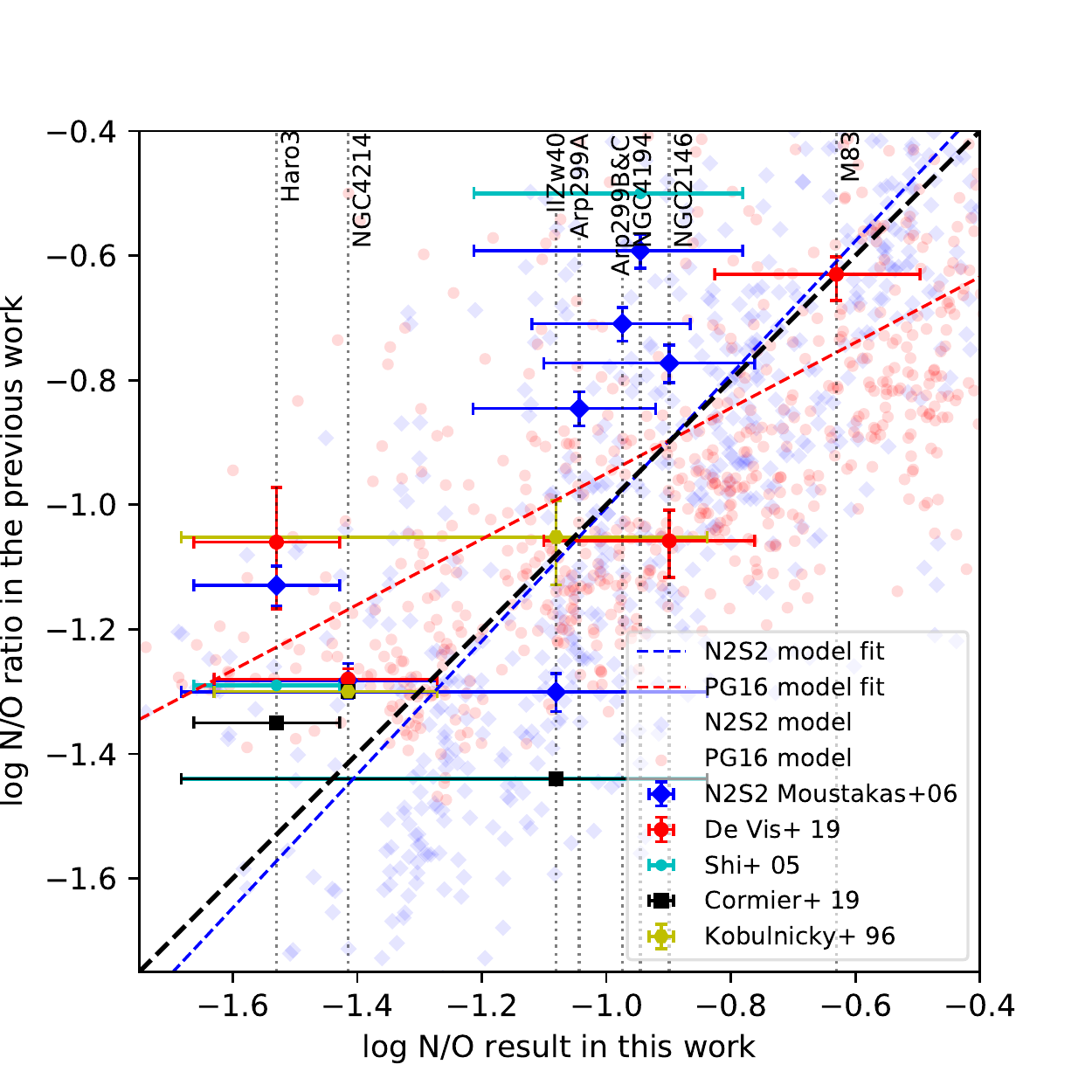}
	\caption{Comparison of the N/O measurement by optical methods to the result of this work. The x-axis is the $\log$ N/O diagnosed through density corrected $N3O3_{n_e}$ index calibrated by Neon line ratio, the y-axis is the N/O measured by optical diagnostics. The opaque data points are plotted as the values listed in Table~\ref{tab:compare}, with different colors and shapes indicating their reference sources. The translucent points are the N/O measurements of photoionization models using $N2S2$ (blue) and PG16 (red) diagnostics. All the vertical dotted lines correspond to the far-IR N/O results for individual galaxies with their names labeled on top. The black dashed line manifests the one-to-one relation. The blue and red dashed line are the least mean square fitting to the blue and red translucent points, representing the relation of the photoionization model N/O ratios measured by the optical $N2S2$ and PG16 methods against the far-IR method.}
\label{f:compare}
\end{figure}

\subsection{Comparison with optical diagnostics}
\label{sec:comparison}

Of the nine galaxies that we have showcased here, only 6 have optical line-based N/O abundance measurements in the literature. 
Haro 3, II Zw 40, and NGC 4214 are in the DGS paper \citep{2019A&A...626A..23C}, while the NGC 4214 N/O measurement is quoted from \cite{1996ApJ...471..211K} based on [N II]$\lambda6584$/[O II]$\lambda3727$ strong line ratio at various positions in long slit observation. 
In addition, \cite{1996ApJ...471..211K} also calculates N/O for II Zw 40. 
Haro 3, M83, NGC 2146 and NGC 4214 are measured in \citet{2019A&A...623A...5D} using a new calibration in \citet{2016MNRAS.457.3678P}, denoted as PG16 hereafter. 
To get a global abundance ratio \citet{2019A&A...623A...5D} combines observations of various scales including fiber, IFU and drift scan, then tries to model the N/O gradient and use the value at $R_{25}$.
One major concern on comparing optical and far-IR N/O measurement is that the optical observations often resolve galaxies and potentially probe smaller regions than the far-IR lines, which use the integrated fluxes of the whole galaxy and measure N/O averaged over a larger aperture.
Hence we calculate N/O using the integrated optical spectroscopic observation in \cite{2006ApJS..164...81M}, and the $N2S2$ index defined in \cite{2009MNRAS.398..949P}. 
This ensures the optical line fluxes also cover the whole galaxy, and the [N II]$\lambda6584$ to [S II]$\lambda6717,6731$ ratio is insensitive to extinction. 
The values taken from these papers and re-calculated with $N2S2$ are listed in Table~\ref{tab:compare} along with those derived from the $N3O3_{n_e}$ model calibration as our best optical estimate of N/O. 
Fig.~\ref{f:compare} shows a comparison of these estimates.

For NGC 4214, NGC 2146 and M83 nucleus, the far-IR derived and optical derived N/O abundance ratios agree within the $1 \sigma$ error. 
But for Arp 299 A and Arp 299 B\&C, the optically derived values are 1.5 to 2 $\sigma$ higher than our $N3O3_{n_e}$ calibrations. 
In the case of Haro 3 and NGC 4194, there are more than one optical measurements. 
For Haro 3, the N/O ratios quoted in \cite{2015A&A...578A..53C} and \cite{2005A&A...437..849S} are higher but within $\sim 2 \sigma$ away from the one-to-one agreement (diagonal line), while the values in the other two sources are more than $4 \sigma$ higher than our far-IR result. 
Similar for NGC 4194, the optical results are higher and at 2 to 2.5 $\sigma$ away from the diagonal line. 
As for II Zw 40, the various optical measurements are off the far-IR estimate but within the $1 \sigma$ errorbar, which is understandable given its highly unreliable measurement of [N\,{\sc iii}] flux.

It is worth noting that the optically derived N/O ratios are overall $\sim$0.2 dex higher than the far-IR results, and N/O measured by different optical probes do not agree with each other. 
Therefore, we also compare the relationship of optical and far-IR N/O diagnostics by computing the N/O ratios of photoionization models using the optical methods against those by our far-IR approach. 
The photoionization models are the ones selected in Sec.\ \ref{sec:intro_model} for consistency. 
We adopted two optical measurements, the $N2S2$ index and PG16, because they are used in our optical N/O calculation and \cite{2019A&A...623A...5D} respectively. 
The optical N/O ratio of models are plotted against far-IR N/O in Fig.\ref{f:compare} as blue ($N2S2$) and red (PG16) translucent points. 
The model N/O ratios calculated by $N2S2$ show a similar trend as the far-IR measurements, and the least mean square fitting relation (shown as the blue dashed line in Fig.~\ref{f:compare}) suggests a close match between the two methods. 
However, the N/O ratio estimated by PG16 systematically deviate from the far-IR method, and follows a different trend as shown by the fitted line (red dashed line) in Fig.~\ref{f:compare}. 
In the figure, PG16 overestimates the N/O ratio in the low N/O regime, then transits to underestimation at N/O beyond -0.9. 
This indicates a significant discrepancy of PG16 calibration with our far-IR method and $N2S2$. 
When compared with the fitted lines, data points of both optical diagnostics show dispersion of 0.4 dex. 
Because $N2S2$ and PG16 are derived through empirical calibration on H {\sc ii} regions, the discrepancy and large scatter seen in the comparison using photoionization models imply systematic differences in the empirical and model based calibrations. 
It might be because either the sample of H {\sc ii} regions used for calibrating these diagnostics don't have enough coverage in the physical parameter space, or the photoionization models explore too large a region in parameter space so that they include physical conditions not present in actual galaxies. 
Calibrations based on individual H {\sc ii} regions could also be at odds with model calibrations that utilize stellar population synthesis, as the latter is more likely to capture the global properties of a galaxy. 
It requires samples of H {\sc ii} regions and photoionization model grids that have finer and wider parameter coverage to further study the relationship and reliability of the two kinds calibrations, but that goes beyond the topic of this work. 
Even though the $N2S2$ shows good agreement with the model calibrated $N3O3_{n_e}$ index, the N/O of our galaxy sample measured by $N2S2$ are still distributed at upper left to the fitted line in Fig.~\ref{f:compare}. 
Hence the difference or the large scatter in the optical to far-IR diagnostics model comparison are not enough to explain the offset in Fig.~\ref{f:compare}. 

There are other factors that may affect the result of far-IR N/O estimate and its comparison to optical techniques. 
The first to consider is the different aperture size for far-IR data. We do not expect beam effect between SOFIA/FIFI-LS and Herschel/PACS for the reasons stated in Sec.\ref{sec:sofia}. 
In the case of NGC 4194, ISO [N\,{\sc iii}] data is used which has a 75 arcsec aperture and could contain more extended flux than in the SOFIA field. 
But that would only contribute to $< 30\%$ [N\,{\sc iii}] flux in ISO data given the small size of NGC 4194, and would lead to overestimation of N/O instead of underestimation. 
As for NGC 2146, it is not affected by aperture size because all the far-IR data are taken from ISO observations. 
Besides, in order to increase the far-IR N/O estimates by 0.2 dex to match with optical results, it would need either $60\%$ more [N\,{\sc iii}] flux, or 36\% less [O\,{\sc iii}]52 flux, or the [Ne\,{\sc iii}]15/[Ne\,{\sc ii}]12 line ratio to increase by at least ten folds. 
The beam effect can not account for such difference. 
Hence we conclude that the aperture size difference of far-IR data can affect the precision of line ratio, but do not cause the lower value of N/O probed by far-IR lines. 

Although the optical N/O results quoted in the previous studies resolve the galaxies, hence they may probe different and much smaller regions than our far-IR measurements, this is not the case for our $N2S2$ calculation which uses spectroscopic observations integrated over the whole galaxies. 
Use of $N2S2$ also avoids the impact of dust extinction. 
So the spatial mismatch of regions and extinction may not be responsible for the discrepancy. 

However, the optical diagnostics may still probe regions different from far-IR in terms of physical conditions. 
Because the optical forbidden lines are enhanced in high temperature regions, while the far-IR fine-structure lines are enhanced in high density regions, we suspect that the optical lines probe hotter and more diffuse ISM than the far-IR lines. 
This means the optical lines are more susceptible to diffuse ionized gas (DIG). 
All the optical abundance probes including PG16 and $N2S2$ rely on the low ionized [N II]$\lambda6584$ line, or other low ionized species such as [O II] and [S II] doublets, for which up to 30\% of the line emission may arise from DIG. 
Although the study by \cite{2017MNRAS.466.3217Z} shows that [N II]/[S II] ratio is less affected by DIG emission than the indexes using Hydrogen recombination lines, we suspect the N/O ratio measured by $N2S2$ is still affected by the abundance in DIG that is not probed by the doubly ionized lines used in $N3O3$. 
This effect can be non-negligible for massive star-forming galaxies, which host most of the recent star formation activities around the nuclei or in a few compact regions, while the DIG across most of the galaxies are enriched by the relatively old population of stars. 
It also indicates an interesting approach to study and dissect the effect of DIG. 

Another factor to consider is that the low and highly ionized lines could come from H {\sc ii} regions with different physical conditions. 
As optical low ionized lines emerge largely from the population of H {\sc ii} regions that host less massive stars with softer radiation, the ISM probed by the low ionized lines have longer life time and could be more enriched with secondary Nitrogen. 
In contrast, far-IR [N\,{\sc iii}] and [O\,{\sc iii}] lines are dominated by the dense ISM surrounding young, massive stars with hard radiation fields. 
These effects combined can also lead to lower N/O ratio probed by highly ionized lines, accounting for the offset shown in the comparison. 
Detailed chemical evolution models combined with photoionization grids would be needed to test this hypothesis. 

We can not yet draw a conclusion for what causes the discrepancy in the optical and far-IR derived N/O ratios. 
But we suggest this could be related to the systematic differences between the empirical and photoionization model based calibration, and that optical and far-IR methods probe ISM of different physical conditions. 
Both questions touch the fundamental question of the reliability and nature of those abundance diagnostics. 
To further validate this new N/O diagnostic in nearby galaxies and study the probable difference in optical and far-IR derived N/O, it is essential to obtain more high-quality [O\,{\sc iii}] 52 $\mu$m and [N\,{\sc iii}]57 $\mu$m spectra with SOFIA/FIFI-LS.

\section{Summary and Prospective Application to High-z Objects}
\label{sec:summary}

In this paper, the far-IR [N\,{\sc iii}]57/[O\,{\sc iii}]52 line ratio is used to define what we call the $N3O3$ parameter. 
We argue it is a robust way to measure the N/O abundance ratio in galaxies because of the co-spatial nature of both ions, and little variation of the emissivity ratio of the lines. 
If the [O\,{\sc iii}] 88 $\mu$m line is available, then $N3O3$ can be corrected for the electron density, which is the primary variable affecting far-IR line ratio. 
This we term the $N3O3_{n_e}$ parameter. 
If the [Ne\,{\sc ii}] 12.8 $\mu$m and [Ne\,{\sc iii}] 15 $\mu$m lines are available, then this diagnostic can be further improved to cope with the deviations of the diagnostic at soft or very hard radiation fields. 
Finally we calibrate the $N3O3$ to Neon line ratio relationship through comparison to photoionization model grids selected from the BOND and CALIFA projects. 
The model calibration first verifies the tight relation of $N3O3$ to derive N/O, showing residual dispersion for all the models to be between 0.05 to 0.1 dex in the range $\log$ [Ne\,{\sc iii}]/[Ne\,{\sc ii}] between -0.5 and 1.5.
It also justifies the need for a density correction, which reduces the dispersion down to only 0.03 dex within a certain range of the Neon line ratio. 
The model calibration in Fig.~\ref{f:n3o3} presents the uncertainty of this diagnostics from the standard deviation in small bins along the x-axis, offering a realistic error estimation for the derived N/O values. 
The model calibration further extends the applicability of $N3O3$ diagnostics to soft and very hard radiation fields.

The $N3O3$ diagnostic is applied to a sample of 9 sources in 8 nearby galaxy systems. 
All the samples are either BCD or LIRGs and host young starburst components. 
The results for deriving the N/O ratio using different levels of calibration show that the original $N3O3$ estimation only differs by $\sim 0.15$ dex for LIRGs when compared with the most precise value output by models that are corrected for both density and ionization effects, while $N3O3_{n_e}$ works better on dwarf galaxies. 
The increased value of the multi-line technique is only achieved for high signal-to-noise line detections. 
In many cases this means that the $N3O3$ or $N3O3_{n_e}$ parameter is sufficient for abundance determinations.

Only six of our sources have optically derived N/O ratios in the literature for comparison, so we also calculate N/O by $N2S2$ index using spectroscopic observations integrated over the whole galaxy, as an attempt to suppress the effect of extinction and spatial mis-match. 
For three sources our far-IR derived N/O ratios are in relatively good agreement with the optically derived values. 
However, for Haro 3, NGC 4194 and both components of Arp 299, the two methods arrive at values that differ by a factor of 1.5 to 2.5, at about $2 \sigma$ significance. 
The optical N/O measurements also appear systematically higher than the far-IR results by $\sim$2 dex. 
To study the relation of different diagnostics, we compared the N/O estimates of photoionization models by those methods. 
We find $N2S2$ agrees well with our model calibrated $N3O3_{n_e}$ index, while PG16 shows large discrepancy, and both of them have dispersion of 0.4 dex. 
We point out that there may be a systematic difference in the empirical and model based calibrations that requires further study.

Though we do not find a definitive explanation for the N/O discrepancy, we can exclude the difference in aperture size, mismatch of observed region and extinction from causing the offset. 
We argue it might be linked to the nature of optical low ionized lines and far-IR [N\,{\sc iii}], [O\,{\sc iii}] lines, with the former probing hotter regions and suffering from DIG contamination, while the latter gives more weights to dense regions surrounding young, massive stars, leading to measurements of N/O ratio of regions of different physical conditions. 
This requires further study with detailed chemical evolution and photoionization modeling. 
The large error from the SOFIA/FIFI-LS spectroscopy which is heavily impacted by telluric absorption may also account for part of the difference. 
To further understand and test this diagnostic, we propose to carry out more observations of [O\,{\sc iii}] and [N\,{\sc iii}] lines in nearby galaxies with SOFIA.

Half of the star formation over cosmic time occurs in dusty star forming galaxies, so the far-infrared methods will be important for extinction free abundance estimates. 
The high-redshift Universe is also surprisingly dusty, so this is where the $N3O3$ diagnostic can achieve its full potential. 
Because it uses just two bright lines that are very close in wavelength, they can be both observed with e.g. the ZEUS-2 instrument \citep{2010SPIE.7741E..0YF} at selected redshift beyond 2 and will shift into ALMA band 10 at $z \sim 5$. 
Since the [O\,{\sc iii}]52 line is among the brightest far-IR lines, it is relatively easy to detect. 
This method could also benefit from the growing number of [O\,{\sc iii}]88 detections at very high redshift \citep[][]{2010ApJ...714L.147F,2016Sci...352.1559I,2020ApJ...896...93H}, as $N3O3_{n_e}$ manifests better performance in the local dwarf galaxies that are thought to resemble high-z galaxies. 
Another advantage is that the high-z galaxies are widely conjectured to have radiation fields that are harder than the local star-forming galaxies \citep{2016ARA&A..54..761S,2019ApJ...882..168P,2020ApJ...896...93H}, pushing the line emission into the regime where the N$^{++}$ and O$^{++}$ species are co-spatial in H {\sc ii} regions. 
This is just the regime where the N/O parameter is least biased and most effective in determining the actual gas phase N/O abundances. 
The N/O can help to solve some of the key questions in the early Universe, including how do galaxies form and grow through star formation across cosmic time; how does the N/O-O/H relationship evolve with time; and what is the relationship between the N/O ratio, metallicity and gas inflow and outflow in the growth of galaxies.

\begin{acknowledgements}

We thank the anonymous referee for the insightful feedback that helped to improve this manuscript.
We thank the SOFIA help desk, and in particular our dedicated support scientist Christian Fisher, for their great help in acquiring and analyzing the SOFIA/FIFI-LS data. 
Support for this work was provided in part by NASA grant NNX171f37G, NSF Grant AST-1716229, NASA/SOFIA grants SOF 04-0179; SOF 05-0111; SOF 06-0225; SOF 07-0209; and SOF 08-0165. 
C. Lamarche acknowledges funding support from NASA grant 80NSSC18K0730.
C. Ferkinhoff acknowledges support by the National Science Foundation under Grant NO. 1847892. 
This work is based on observations made with the NASA/DLR Stratospheric Observatory for Infrared Astronomy (SOFIA). SOFIA is jointly operated by the Universities Space Research Association, Inc. (USRA), under NASA contract NNA17BF53C, and the Deutsches SOFIA Institut (DSI) under DLR contract 50 OK 0901 to the University of Stuttgart. 
Herschel is an ESA space observatory with science instruments provided by European-led Principal Investigator consortia and with important participation from NASA. 
The Infrared Space Observatory (ISO) is an European Space Agency (ESA) mission with the participation of ISAS (Japan) and NASA (USA). This Web server is maintained at the ISO Data Centre, which is based at Villafranca, Madrid, and is part of the Science Operations and Data Systems Division of the Research and Scientific Support Department. 
This work is based in part on archival data obtained with the Spitzer Space Telescope, which was operated by the Jet Propulsion Laboratory, California Institute of Technology under a contract with NASA.

\end{acknowledgements}


\bibliography{main}


\clearpage


\appendix

\section{Calculation of $N3O3$ and density correction}
\label{sec:calculation}

\subsection{Basic of detailed balancing}
\label{sec:coefficient}

The atomic parameters governing collisional excitation and emission of the [N\,{\sc iii}] and [O\,{\sc iii}] far-IR fine-structure lines are summarized in Table~\ref{tab:coefficient}. 
The collisional strengths between the levels of the ground state term of O$^{++}$ and N$^{++}$ are taken from \citet{2017ApJ...850..147T} and \citet{1992ApJS...80..425B}, both evaluated at $T_e=10^4 \ \mathrm{K}$. 

\begin{table}[h]
\centering
\begin{tabular}{llllll}
	\hline
	 & \multicolumn{1}{c}{N\,{\sc iii}} &\ & \multicolumn{3}{c}{O\,{\sc iii}} \\ \cline{2-2} \cline{4-6}
	Transition & $^2P_{1/2}, ^2P_{3/2}$ & & $^3P_0, ^3P_1$ & $^3P_2, ^3P_1$ & $^3P_0, ^3P_2$ \\ \hline
	$\lambda$($\mu$m) & 57.32 & & 88.36 & 51.81 & 32.7 \\
	$\Omega(i, j)$ & 1.445 & & 0.542 & 1.28 & 0.261 \\
	$A_{ij}$ & $4.8 \times 10^{-5}$ & & $2.6\times10^{-5}$ & $9.8 \times 10^{-5}$ & $3.0 \times 10^{-11}$ \\ \hline
\end{tabular}
\caption{The collisional and emission parameters used for the emissivity calculation of the [N\,{\sc iii}] and [O\,{\sc iii}] emitting levels. The rows are the transition associated with the line, line wavelength $\lambda$, collisional strengths $\Omega(i, j)$ between energy level $i$ and $j$, and the spontaneous emission coefficient $A_{i, j}$}
\label{tab:coefficient}
\end{table}
The collisional excitation rate follows the definition in \citet{1989agna.book.....O}. 
Here we ignore the electron velocity (temperature) dependence of the rate coefffient. 
We also assume the line emission is optically thin: self-absorption and and stimulated emission are not important.

\subsection{$N3O3$ at low density limit}
\label{sec:lowdensity}

In the low density limit ($n_{e} \ll n_{crit}$) approximation, collisional de-excitation is unimportant, and all the ions at the excited states will eventually transit downwards by emitting a photon. 
So only collisional excitation and radiative de-excitation are considered. 
Since the low density limit means radiative transitions are much faster than collisional transitions, to a good approximation, all ions are in the ground state so that the total populations of the ions equals the ground state population. 
For the [N\,{\sc iii}] line in two level system, the low density limit balance is reached when $n_{\mathrm{tot}} n_{e} q_{01} = n_1 A_{10}$, where $q_{01}$ denotes the collisional coefficient from ground state 0 to the first excited state 1. 
The emissivity is defined here as the power emitted in the specific line per particle per unit time, ignoring the solid angle factor for simplicity. 
It is calculated as
\begin{equation}\begin{split}
\label{equ:niii_lowdensity}
	\varepsilon_{\mathrm{[N\,III]},n_e\rightarrow0} = h\nu_{\mathrm{[N\,III]}}\ A_{10} \frac{n_1}{n_{\mathrm{tot}}} = 2.161 \times 10^{-19}\ \frac{n_e}{T_e^{1/2}} \ \mathrm{erg\ s^{-1}} 
\end{split}\end{equation}

The calculation for the [O\,{\sc iii}] 52 $\mu$m line follows the same logic, but the balance is between collisional excitation from $^3P_0$ to $^3P_2$, and radiative transitions down from $^3P_2$ to $^3P_1$. 
Line emission between the $^3P_2$ and $^3P_0$ states is highly forbidden and safely ignored. 
Therefore, the emissivity of the [O\,{\sc iii}]52 line in the low density limit is
\begin{equation}\begin{split}
\label{equ:oiii_lowdensity}
	\varepsilon_{\mathrm{[O\,III]52},n_e\rightarrow0} = 8.635 \times 10^{-20}\ \frac{n_e}{T_e^{1/2}} \ \mathrm{erg\ s^{-1}} 
\end{split}\end{equation}

Then the N3O3 is defined as the line flux ratio divided by the ratio of emissivity, 
\begin{equation}\begin{split}
	\frac{N}{O} \ \sim \frac{F(\mathrm{[N\,III]})}{F(\mathrm{[O\,III]}52)}\ \frac{\varepsilon_{\mathrm{[O\,III]52,n_e\rightarrow0}}}{\varepsilon_{\mathrm{[N\,III]},n_e\rightarrow0}} = N3O3
\end{split}\end{equation}

\subsection{$N3O3$ with $n_e$ dependence}
\label{sec:densitycorr}

To include the density dependence of the line ratios, the balance of collisional excitation/de-excitation and spontaneous radiation from all the levels in the ground state term must be taken into account. 
The optically emitting excited terms in the electronic configuration lie high ($\sim$ 30,000 K) above ground and have large radiative transition probabilities so their level populations will be small compared with those in the ground term. 
Ignoring these transitions will therefore have negligible effects on far-IR line emission.

For the N$^{++}$ ion, the detailed balancing is then $n_e n_1 q_{10} + n_1 A_{10} = n_e n_0 q_{01}$ and the emissivity is 
\begin{equation}\begin{split}
	\varepsilon_{\mathrm{[N\,III]}} = h\nu_{\mathrm{[N\,III]}}\ A_{10} \frac{n_e q_{01}}{n_e (q_{01} + q_{10}) + A_{10}} = \frac{2.161 \times 10^{-19} }{0.1949 + \tfrac{T_e^{1/2}}{n_e}} \ \mathrm{erg \ s^{-1}}
\end{split}\end{equation}

The calculation for the O$^{++}$ ion is similar but more complicated as it involves 3 levels. 
The equations are straightforward to solve \citep[c.f.][]{1989agna.book.....O} and the net result is:
\begin{equation}\begin{split}
	\varepsilon_{\mathrm{[O\,III]52}} = h\nu_{\mathrm{[O\,III]52}} A_{21} \frac{n_2}{n_\mathrm{tot}} = 8.635 \frac{\tfrac{n_e}{T_e^{1/2}}\left(1+ 0.4956\tfrac{n_e}{T_e^{1/2}}\right)}{1+0.3766 \tfrac{n_e}{T_e^{1/2}} + 0.02050\left(\tfrac{n_e}{T_e^{1/2}}\right)^2} \times 10^{-20} \ \mathrm{erg\ s^{-1}}
\end{split}\end{equation}

In the low density limit, the line emissivity approaches the result calculated above Equ.~\ref{equ:niii_lowdensity} and ~\ref{equ:oiii_lowdensity}. 
The density corrected line ratio is then expressed as:
\begin{equation}\begin{split}
\label{equ:n3o3_ne_emissity}
	\frac{N}{O}\sim \frac{F(\mathrm{[N\,III]})}{F(\mathrm{[O\,III]}52)}\ \frac{\varepsilon_{\mathrm{[O\,III]52}}}{\varepsilon_{\mathrm{[N\,III]}}} = N3O3 \times \frac{\left(0.1621 + \tfrac{T_e^{1/2}}{n_e}\right)\tfrac{n_e}{T_e^{1/2}}\left(1+ 0.4956\tfrac{n_e}{T_e^{1/2}}\right)}{1+0.3766 \tfrac{n_e}{T_e^{1/2}} + 0.02050\left(\tfrac{n_e}{T_e^{1/2}}\right)^2} = N3O3_{n_e}
\end{split}\end{equation}

If we then introduce the temperature dependence $\tfrac{n_e}{T_e^{1/2}}$ factor in Equ.~\ref{equ:n3o3_ne_emissity} with the [O\,{\sc iii}]52/88 line ratio $R_{52/88} = 10.72 - \frac{396.4}{\frac{n_e}{T_e^{1/2}} + 39.00}$, we get the diagnostic as a function of [O\,{\sc iii}]52/88 in Equ.~\ref{equ:n3o3_ne_r21}.

\section{SOFIA/FIFI-LS data}
\label{sec:sofia_data}

Each SOFIA/FIFI-LS observation is represented with a panel of 2 figures, the first one is the moment 0 map of the spectral line with an ellipse showing the region over which the spectrum is measured, 
the first figure is the corresponding continuum subtracted spectrum. The line flux map is color coded by the flux intensity of each pixel shown as the colorbar. The solid contours correspond to +3, +6, +9, ... times the median of error per pixel in the field. The dashed line labels 0 flux, and dotted lines are -3, -6 the median error levels. In the second figure, the spectrum and 1-sigma error is plotted as black solid line, with the raw atmospheric transmission at the time of observation and that smoothed to the FIFI-LS spectral resolution presented in grey dotted and solid lines. The channels that are integrated to make the line flux map and to measure the line flux are shown in yellow.

\begin{figure}[h]
	\includegraphics[width=\textwidth]{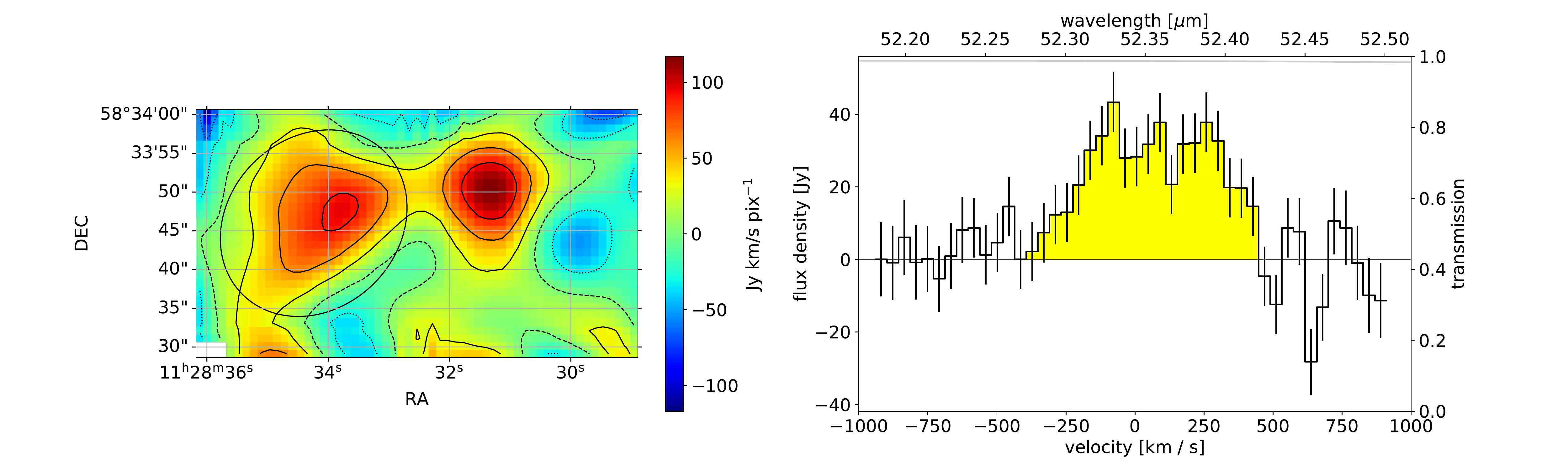}
	\label{f:Arp299a_oiii}
    \caption{Arp 299 A [O\,{\sc iii}]52 line map and spectrum. For details see text in Appendix~\ref{sec:sofia_data}.}
\end{figure}

\begin{figure}[h]
	\includegraphics[width=\textwidth]{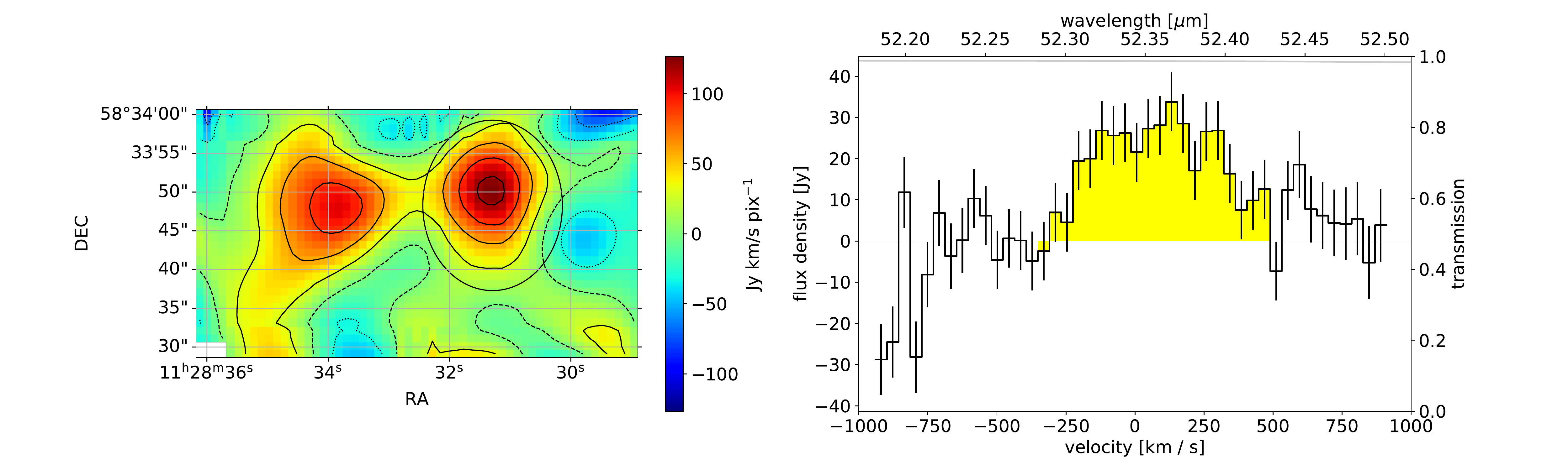}
	\label{f:Arp299bc_oiii}
    \caption{Arp 299 B\&C [O\,{\sc iii}]52 line map and spectrum. For details see text in Appendix~\ref{sec:sofia_data}.}
\end{figure}

\begin{figure}[h]
	\includegraphics[width=\textwidth]{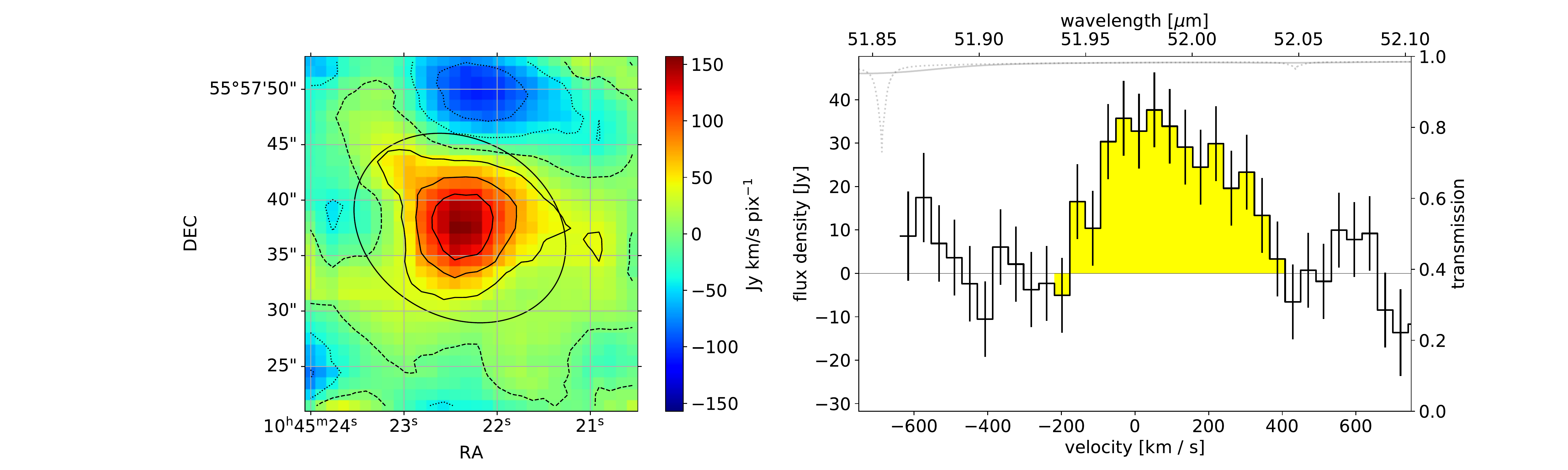}
	\label{f:haro3_oiii}
    \caption{Haro 3 [O\,{\sc iii}]52 line map and spectrum. For details see text in Appendix~\ref{sec:sofia_data}.}

\end{figure}

\begin{figure}[h]
	\includegraphics[width=\textwidth]{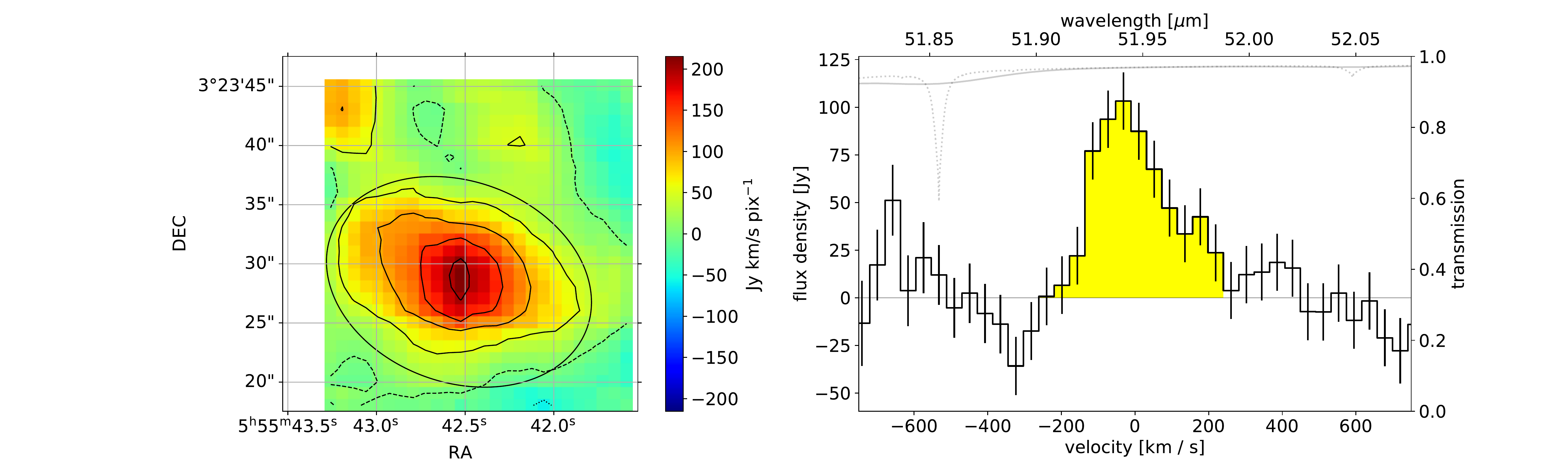}
	\label{f:iizw40_oiii}
    \caption{II-Zw 40 [O\,{\sc iii}]52 line map and spectrum. For details see text in Appendix~\ref{sec:sofia_data}.}

\end{figure}

\begin{figure}[h]
	\includegraphics[width=\textwidth]{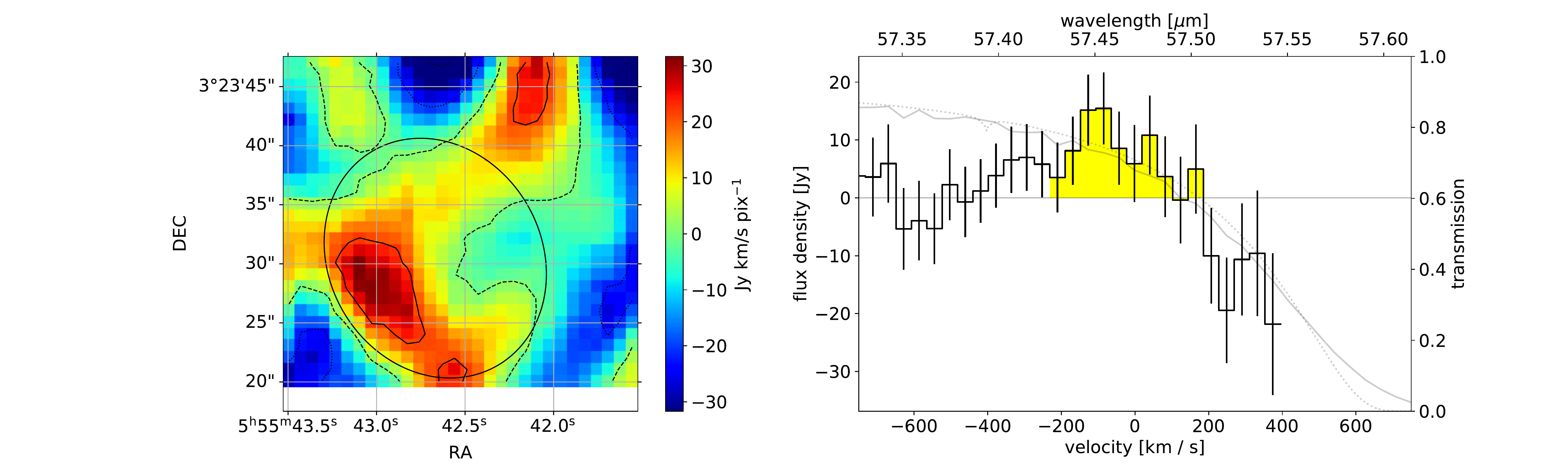}
	\label{f:iizw40_niii}
    \caption{II-Zw 40 [N\,{\sc iii}] line map and spectrum. For details see text in Appendix~\ref{sec:sofia_data}.}

\end{figure}

\begin{figure}[h]
    \includegraphics[width=\textwidth]{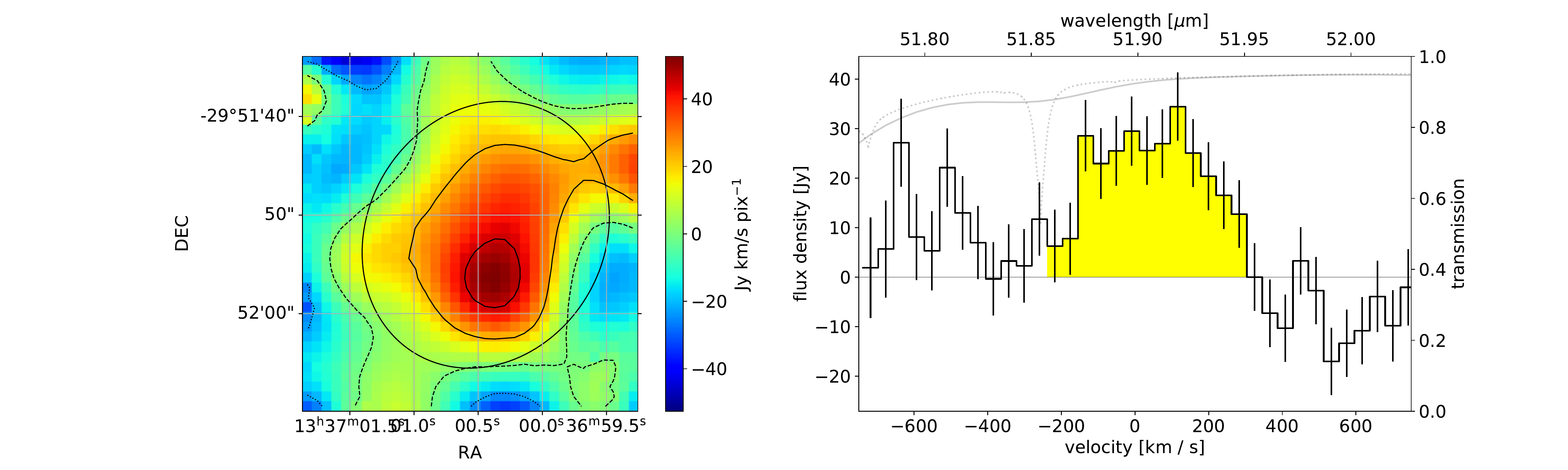}
    \label{f:m83_oiii}
    \caption{M 83 nucleus [O\,{\sc iii}]52 line map and spectrum. For details see text in Appendix~\ref{sec:sofia_data}.}

\end{figure}

\begin{figure}[h]
	\includegraphics[width=\textwidth]{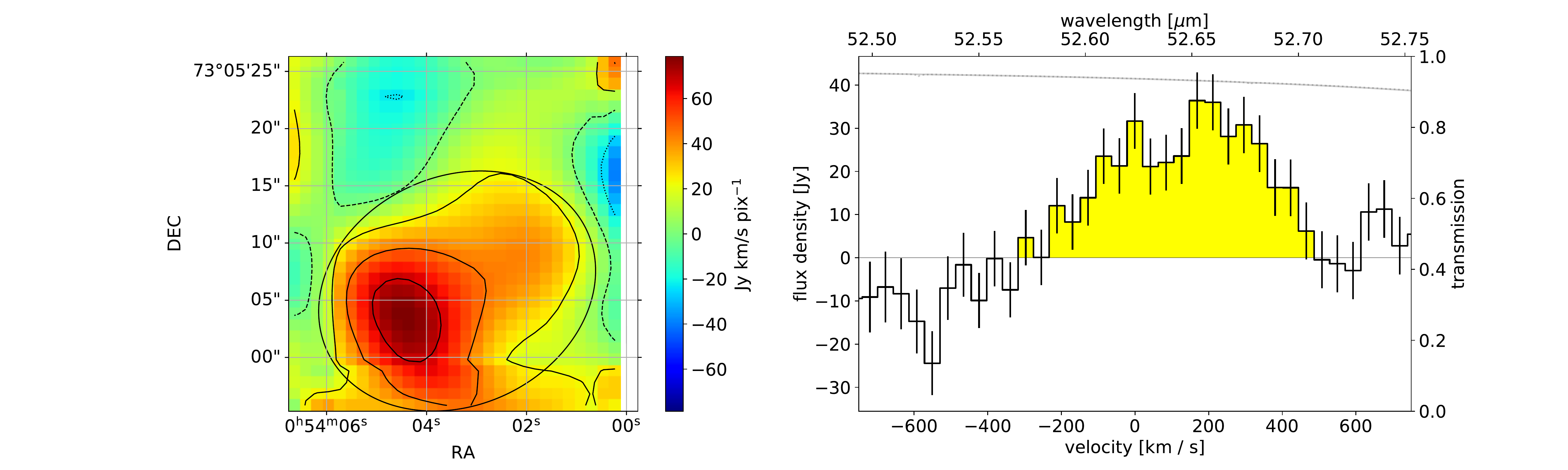}
	\label{f:mcg+12_oiii}
	\caption{MCG+12-02-001 [O\,{\sc iii}]52 line map and spectrum. For details see text in Appendix~\ref{sec:sofia_data}.}
\end{figure}

\begin{figure}[h]
	\includegraphics[width=\textwidth]{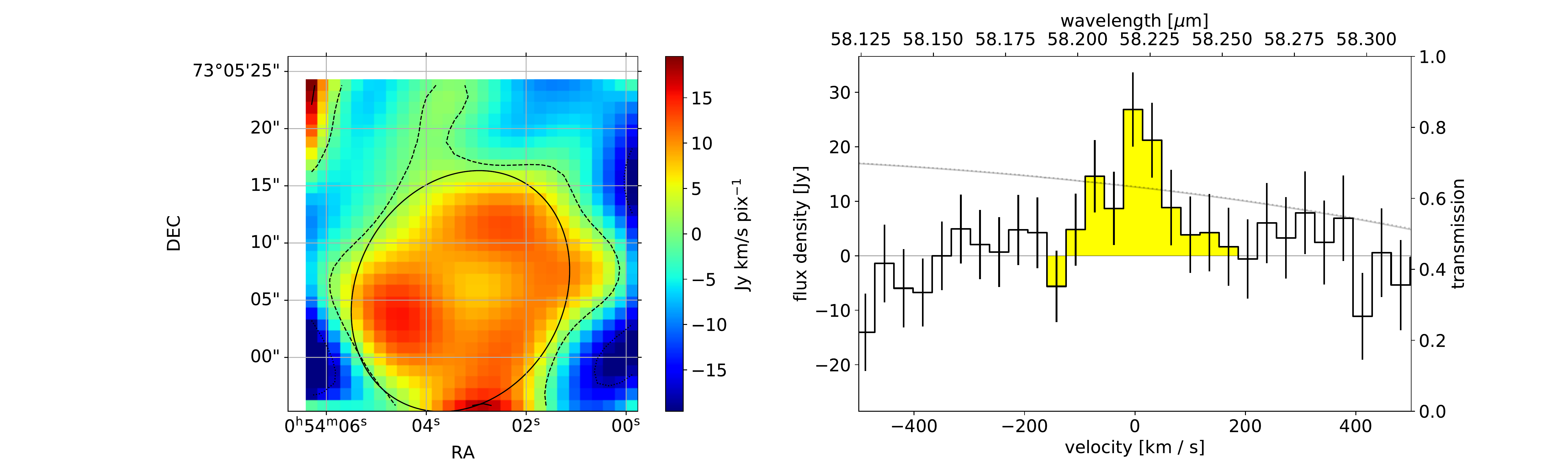}
	\label{f:mcg+12_niii}
	\caption{MCG+12-02-001 [N\,{\sc iii}] line map and spectrum. For details see text in Appendix~\ref{sec:sofia_data}.}

\end{figure}

\begin{figure}[h]
	\includegraphics[width=\textwidth]{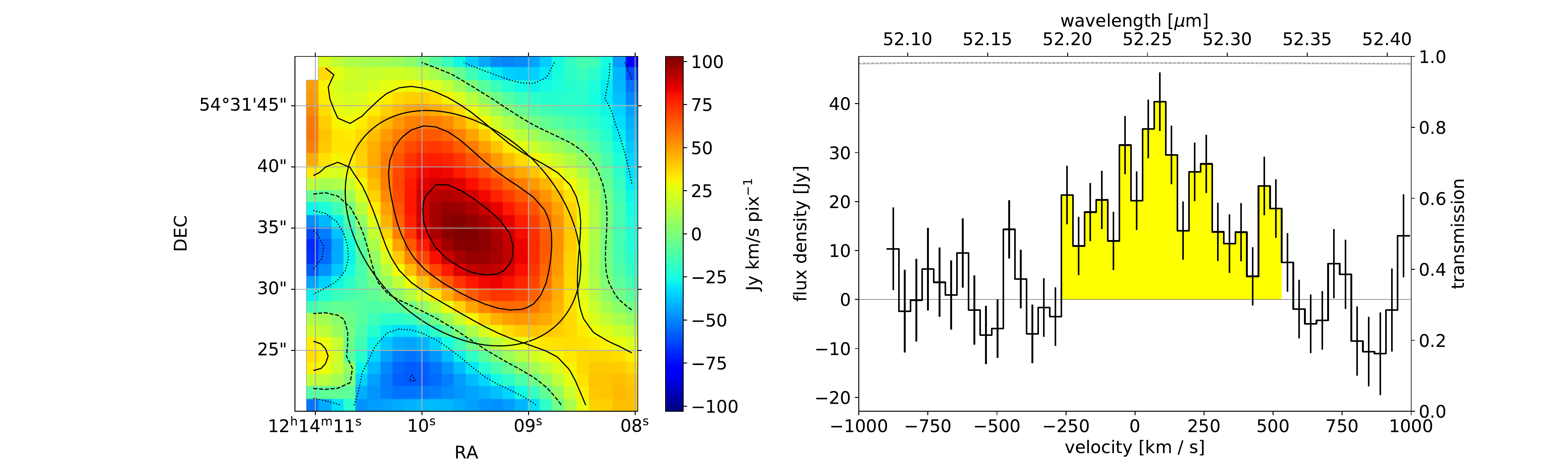}
	\label{f:ngc4194_oiii}
	\caption{NGC 4194 [O\,{\sc iii}]52 line map and spectrum. For details see text in Appendix~\ref{sec:sofia_data}.}

\end{figure}

\begin{figure}[h]
	\includegraphics[width=\textwidth]{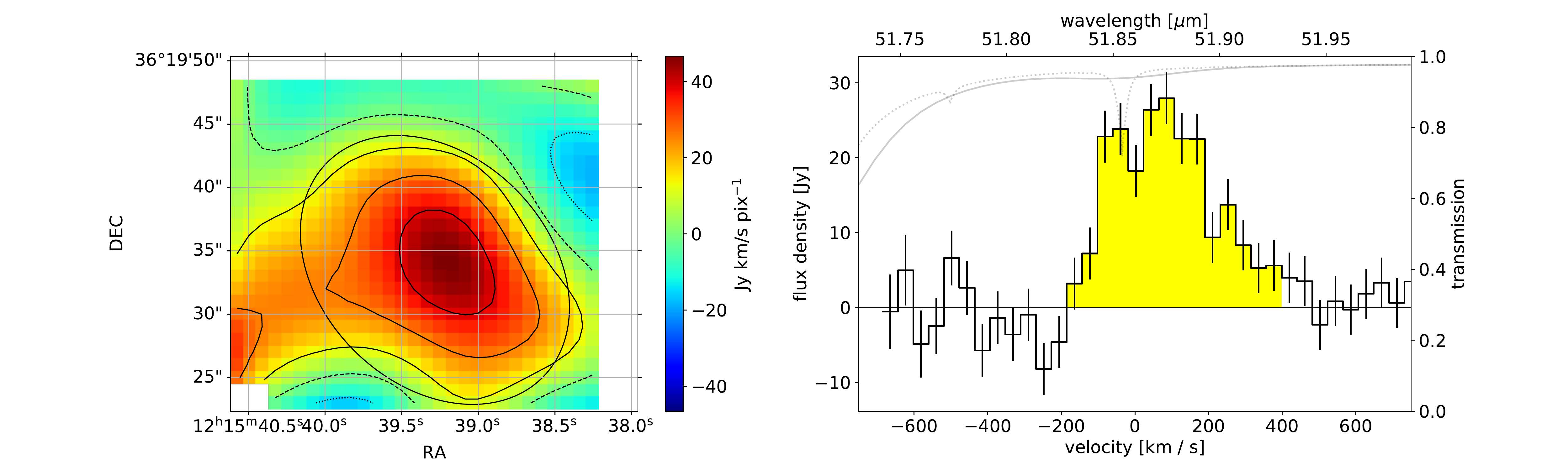}
	\label{f:ngc4214_oiii}
	\caption{NGC 4214 region I [O\,{\sc iii}]52 line map and spectrum. For details see text in Appendix~\ref{sec:sofia_data}.}

\end{figure}

\begin{figure}[h]
	\includegraphics[width=\textwidth]{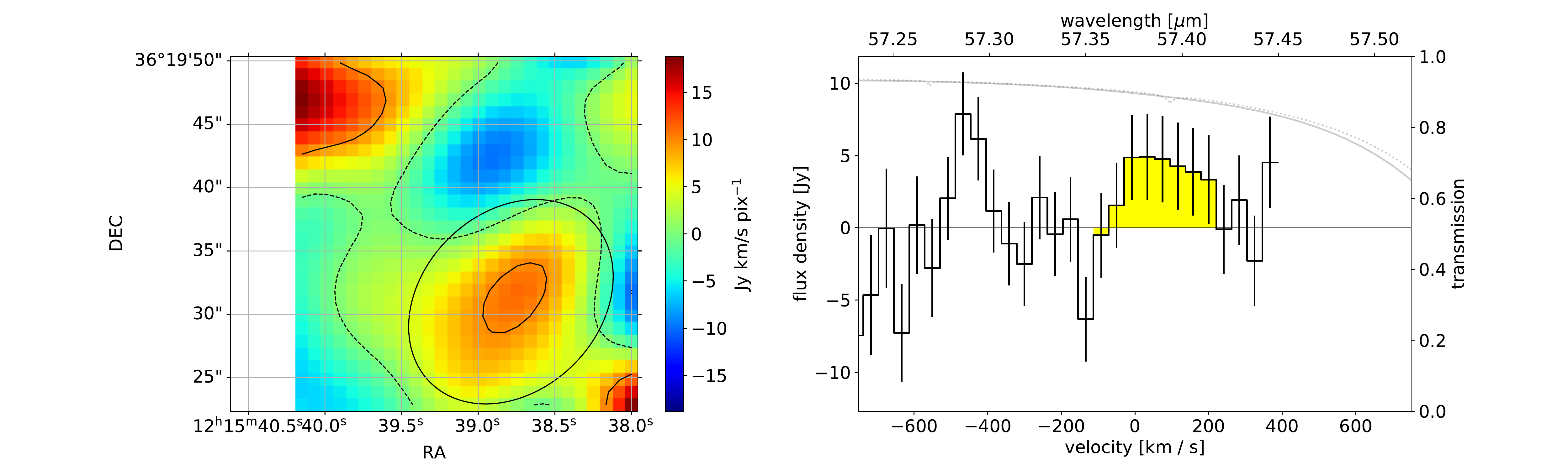}
	\label{f:ngc4214_niii}
	\caption{NGC 4214 region I [N\,{\sc iii}] line map and spectrum. For details see text in Appendix~\ref{sec:sofia_data}.}
\end{figure}

\end{document}